\documentclass{jfm}

\usepackage{amsmath}
\usepackage{color}
\usepackage{xfrac}

\usepackage{graphicx}
\graphicspath{{figures/}}

\usepackage{amssymb}
\usepackage{amsbsy}
\usepackage{upmath}
\usepackage{textcomp}
\usepackage{natbib}

\newcommand{\RomanNumeralCaps}[1]
\linenumbers

\def \ubu {\boldsymbol{u}}
\def \ustar {\ubu^\ast}
\def \Tstar {\ubtau^\ast}

\shorttitle{Flow onset for a single bubble in yield-stress fluids}
\shortauthor{A.~Pourzahedi et al.}

\title{Flow onset for a single bubble in yield-stress fluids}

\author{Ali Pourzahedi\aff{1},
Emad Chaparian\aff{2}
\corresp{\email{emad@math.ubc.ca}},
Ali Roustaei\aff{3}
\corresp{\email{aroustaei@ut.ac.ir}}
\and\\
Ian A.~Frigaard\aff{1,2}}

\affiliation{
\aff{1}Department of Mechanical Engineering, University of British Columbia, 2054-6250 Applied Science Lane, Vancouver, BC, Canada V6T 1Z4
\aff{2}Department of Mathematics, University of British Columbia, 1984 Mathematics Road, Vancouver, BC, Canada, V6T 1Z2
\aff{3}School of Engineering Science, College of Engineering, University of Tehran, Tehran, Iran
}

\begin{document}
\maketitle

\begin{abstract}
We use computational methods to determine the minimal yield-stress required in order to hold static a buoyant bubble in a yield-stress liquid. The static limit is governed by the bubble shape, the dimensionless surface tension ($\gamma$) and the ratio of the yield-stress to the buoyancy stress ($Y$). For a given geometry, bubbles are static for $Y > Y_c$, which we determine for a range of shapes. Given that surface tension is negligible, long prolate bubbles require larger yield-stress to hold static compared to oblate bubbles. Non-zero $\gamma$ increases $Y_c$ and for large $\gamma$ the yield-capillary number ($Y/\gamma$) determines the static boundary. In  this limit, although bubble shape is important, bubble orientation is not. 2D planar and axisymmetric bubbles are studied.
\end{abstract}

\begin{keywords}
Multiphase flow, Bubble dynamics, Plastic materials
\end{keywords}

\section{Introduction}
\label{sec:intro}

\begin{figure}
\centerline{\includegraphics[width=0.9\linewidth]{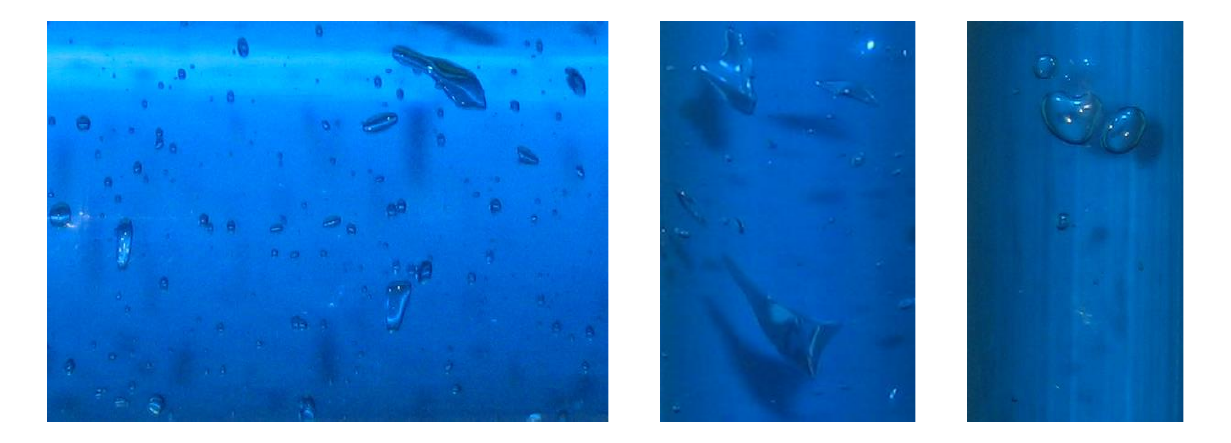}}
\caption{A range of static bubbles in a Carbopol gel (similar effects with many commercial hair gels).}
\label{fig:bubbles}
\end{figure}

With a closed tube of a hydrogel and some vigorous shaking one can create a wide range of stationary bubbles; see figure \ref{fig:bubbles}. These range in size over many decades. The smaller ones may be more spherical or elliptic, presumably as surface tension is significant. Larger bubbles can be quite angular and contain concavities. There is no obvious orientation with respect to gravity. The ability of the fluid to resist both buoyancy and surface tension forces and remain stationary, necessitates a constitutive law with a finite (deviatoric) stress at zero strain rate, i.e.~a yield-stress or equivalent. This is a purely dimensional argument. The simplest such fluids are described by viscoplastic models such as the Bingham fluid.

Viscoplastic fluids are ubiquitous in a wide variety of industrial and medical applications, as well as geophysical sciences. The common characteristic of all these materials is their yield-stress, meaning that they flow only if the stress applied is greater than a yield value; otherwise they exhibit a solid-like behavior \citep{balmforth2014yielding}. The dynamics of bubbles in viscoplastic fluids contain many interesting problems that are not fully explored. In this study we focus on the boundary between flow and entrapment, for single bubbles in a yield-stress fluid. As figure \ref{fig:bubbles} suggests, shape, size and surface tension effects are all of importance and will be explored.

More than a toy problem, static bubbles in viscoplastic fluids occur in many industrial settings. In the oil sands industry, the by-products of the extraction of bitumen from oil sands are stored in tailings ponds over many decades. Pond liquids, known as “Fluid Fine Tailings” and “Mature Fine Tailings” (FFT/MFT) are complex suspensions rheologically characterised as thixotropic yield-stress fluids \citep{derakhshandeh2016kaolinite}. Anaerobic micro-organisms within the fluid form both carbon dioxide and methane. Management of emissions presents a difficult environmental challenge to the oil sands industry \citep{small2015emissions}, and there is consequent interest in estimating the degree to which FFT/MFT can retain gas bubbles. Similar phenomena occur in the nuclear waste slurry, where flammable (Hydrogen) gas rises from the viscoplastic waste suspension \citep{johnson2017yield}. Detailed studies of gas bubbles in nuclear waste tanks, which also contain thixotropic yield-stress material, show a wide range of shapes in retained bubbles \citep{gauglitz1996mechanisms}. Similar mechanisms occur in geological materials, such as shallow marine, terrestrial sediments and some flooded soils, giving a wider relevance to questions of bubble formation and release \citep{boudreau2012physics}. In the food industry, entrapment of air bubbles inside products may give them a different texture and flavor, such as aerated chocolate. It may also affect the efficiency of fermentation processes. In the cosmetic industry, products such as hair gel are often sold by volume, bubbles included.

While the above are concerned with bubble entrapment in stationary fluids, the phenomenon is also of interest in flowing fluid. In the oil and gas industry, influx of formation gases into the drilling mud (generally a viscoplastic fluid) might occur during well construction, known as gas kick. If uncontrolled, the gas will rise to the surface, potentially causing a blowout with severe safety and environmental hazards. When controlling kicks the (gas cut) drilling mud is circulated slowly from the well, which is closed in and under pressure. Whereas large gas bubbles generally rise through the mud, smaller bubbles may remain trapped \citep{johnson1991gas,johnson1995gas}. The trapped gas fraction is however hazardous to well control operations \citep{Gonzalez2000} and needs to be accounted for in operations. It is worth commenting that the trapping of gas by the yield-stress of a fluid is analogous to the retention of dissolved gas under pressure e.g.~carbon dioxide in deep lakes. Risks associated with these configurations (i.e.~limnic eruptions), thus have yield-stress analogues for the industrial processes discussed.

Despite the interest in these problems, there are still relatively few studies. The motion of bubbles in a viscoplastic fluid has been studied theoretically and numerically in the literature by \citet{dubash2004, tsamopoulos2008, dimakopoulos2013steady, singh2008interacting, chaparian2021clouds, tripathi2015bubble, karapetsas2019dynamics, deoclecio2021bubble}.
\citet{tsamopoulos2008} comprehensively studied the steady rise of a single axisymmetric bubble in a viscoplasic fluid. They investigated the effect of inertia, surface tension and yield-stress on the bubble shape and velocity. Later, \citet{dimakopoulos2013steady} expanded the problem to Herschel--Bulkley fluids and compared the previous regularization approach with the augmented Lagrangian method, finding relatively similar results.
Interaction of multiple rising bubbles or falling droplets in a Bingham fluid is investigated in 2D by \citet{singh2008interacting}, again using a regularization method. They observed that multiple bubbles with the same size, aligned vertically and close to each other, can overcome the yield-stress, where a single bubble would remain trapped. In \citet{chaparian2021clouds} we have studied the effects of multiple bubbles (clouds/suspensions) on the static stability limit using a Monte-Carlo approach, finding that the interaction of clusters of bubbles is all important in determining the flow onset.
\citet{tripathi2015bubble} studied the rise of axisymmetric bubbles in a Bingham fluid using a regularization method. They have shown that in the presence of inertia, high yield-stress and low surface tension can lead to an unsteady oscillating rise of bubbles.
\citet{karapetsas2019dynamics} have investigated the bubble rise dynamics when subjected to an acoustic pressure field and found that acoustic excitation accelerates the motion of the bubble by increasing the size of the yielded region surrounding the bubble. In a recent study, \citet{deoclecio2021bubble} have analyzed the entrapment conditions of initially spherical and ellipsoidal bubbles in a regularized Bingham fluid. They have found that surface tension does not play a role in entrapment of spherical bubbles, however, it will facilitate the rise of non-spherical bubbles by yielding of the surrounding fluid.

While in Newtonian fluids, the Stokes flow limit gives a slowly moving spherical bubble, the Stokes flow limit of a bubble in a yield-stress fluid is non-unique for static configurations. \citet{tsamopoulos2008} study these scenarios as a limit of steadily moving bubbles. For different Bond numbers the flow stops at different critical yield-stresses. However, their procedure also iterates to find the shape of the steadily moving bubble: they take spherical bubble as the initial shape only. Thus, shape effects are not studied independently in their formulation. As figure \ref{fig:bubbles} shows, if the static problem is studied without limiting from a moving bubble, there is little restriction on the bubble shape that can be trapped. This explains our approach here, in which we control the shape and study the critical limit for flow onset. This non-uniqueness and the general theoretical framework for static bubbles was first expounded by \citet{dubash2004}, who formally defined the critical Bingham (yield) number above which a given bubble shape remains trapped in a viscoplastic fluid, using the variational methods. Here we explore this approach computationally.

In addition to the numerical and analytical methods, there have been several experimental studies trying to better understand buoyancy-driven rise of bubbles in viscoplastic materials \citep{astarita1965motion, terasaka2001bubble,dubash2007propagation, sikorski2009motion, mougin2012significant, zare2021effects, lopez2018rising, pourzahedi2021eliminating}. The first notable work in this field was performed by \cite{astarita1965motion}, where they reported the shape as well as the correlation between velocity and volume of gas bubbles in various non-Newtonian fluids. \citet{terasaka2001bubble} experimentally investigated the effect of operating parameters such as injection nozzle diameter, gas flow rate and rheological parameters on the bubble volume released into viscous yield-stress fluids. \citet{dubash2007propagation} as well as \citet{sikorski2009motion} performed a similar experimental study, and used the results to predict the stopping condition for the rise of single bubbles using an energy budget model. In general such models have not been successful in that they extrapolate from the behavior of moving bubbles, which are typically far from the critical conditions, may have different shape, etc.
In another study, \citet{mougin2012significant} performed Particle Image Velocimetry (PIV) analysis in order to find the local velocity and strain fields. They investigated the influence of internal stresses existing within the yield-stress fluids on the trajectory and shape of bubbles. Recently, \citet{zare2021effects} have further explored the effect of `damaged' pathways created by previous bubbles on the trajectory of the subsequent bubbles.
\citet{lopez2018rising}, studied the effect of yield-stress, inertia, buoyancy and elasticity on the bubble shape and velocity using various concentrations of a Carbopol solution. An interesting feature of experimentally observed bubbles \citep{dubash2007propagation,sikorski2009motion,lopez2018rising, pourzahedi2021eliminating}, is that they tend to adopt an inverted teardrop shape as they rise. This is quite different to computational results and it has been suggested that the root cause is viscoelasticity, causing the tail to extend \citep{tsamopoulos2008}.

In this study we fill the knowledge gap of quantifying the limits for which bubbles rise or remain trapped. While we cannot cover all possible shapes, we study families of bubble shapes so that the characteristic effects of aspect ratio and curvature can be understood. For these families we explore the relative effects of surface tension and buoyancy on flow onset. Although we also compute non-zero flows, the emphasis is on the static limit. The definition of the problem and its governing equations is given in \S \ref{sec:formulation}. The computational methods are presented in \S \ref{sec:methods}. We compare results for elliptical bubbles with the slipline theory (perfect plasticity), as a pseudo-benchmark. The effects of aspect ratio and surface tension on the critical yield number for 2D elliptical and quartic bubbles are explored in \S \ref{sec:Results}.
Axisymmetric results are presented in \S \ref{sec:ResultsAx}. We study families of  elliptic bubbles, where the results are qualitatively similar to the planar 2D bubbles. We also explore inverted tear drop shaped bubbles similar to those found experimentally, far from the yield limit. We close the study with some concluding remarks regarding the findings and future directions.

\section{Formulation}
\label{sec:formulation}

\begin{figure}
\centerline{\includegraphics[width=0.3\linewidth]{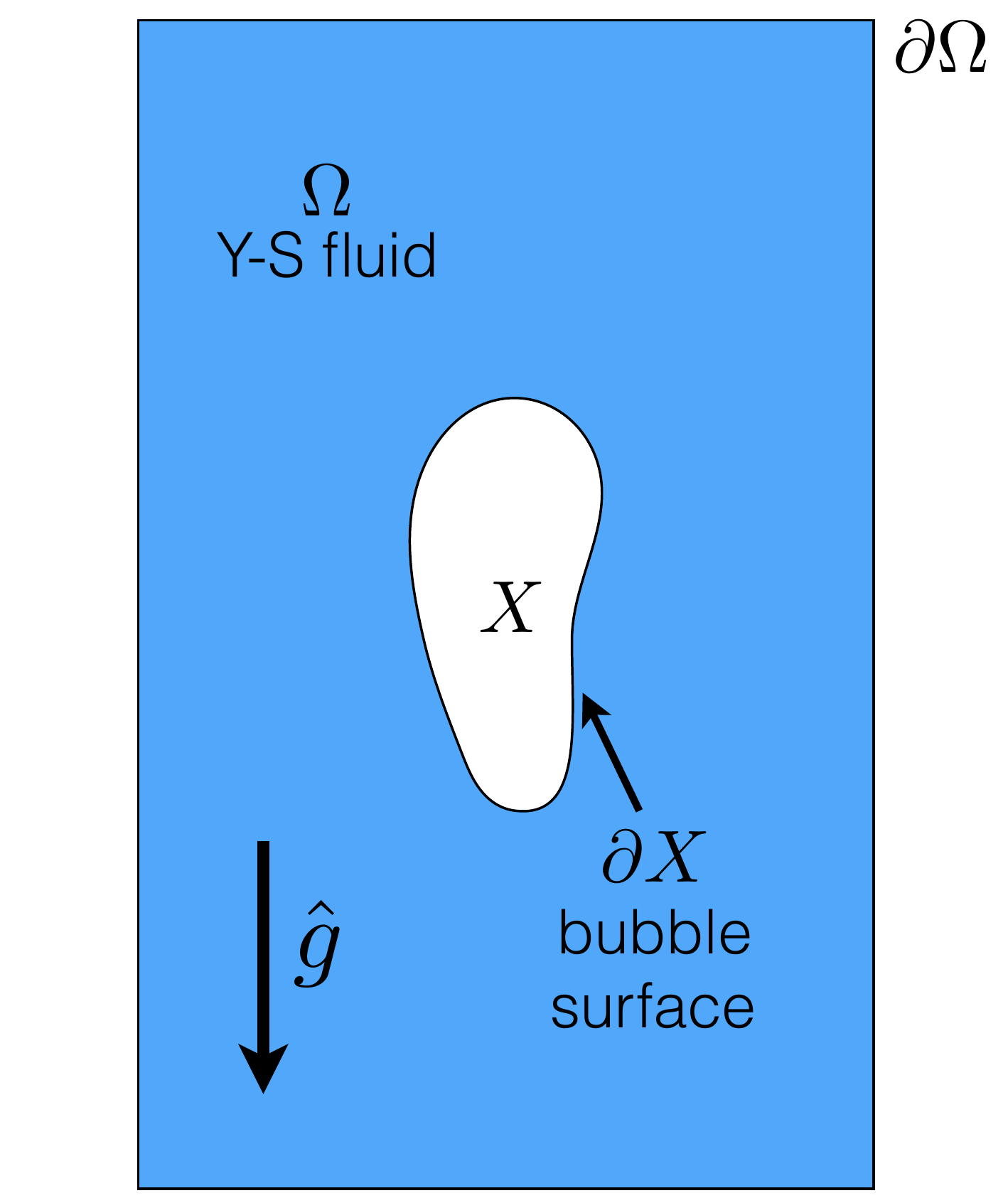}}
\caption{Schematic of the flow setup.}
\label{fig:schematic}
\end{figure}

The general setting considered is that of a single bubble in an expanse of yield-stress fluid; see figure \ref{fig:schematic}. We mostly consider the yield limit of bubbles, which is the same for all yield-stress fluids using a von Mises yield criterion, and hence we consider the simplest Bingham fluid model. Similarly, due to the focus on the yield limit, we may consider the flow to be incompressible and non-inertial. Throughout this paper, the $\hat{\cdot}$ accent signifies a dimensional parameter or variable.
\begin{eqnarray}\label{govern}
0 &=& - \boldsymbol{\nabla} \hat{p} + \boldsymbol{\nabla} \boldsymbol{\cdot} \hat{\ubtau} - \hat{\rho}_f \hat{g} ~\boldsymbol{e}_g ~~\text{in}~~ \Omega \setminus \bar{X}, \\
0 &=&  \boldsymbol{\nabla} \boldsymbol{\cdot} \hat{\boldsymbol{u}}.
\end{eqnarray}
The bubble is finite and the flow is driven by the buoyancy of the bubble. Thus, distant from the bubble, the stress falls below the yield-stress of the fluid and the flow is stagnant. Without loss of generality we impose zero velocity on the flow in the far-field.

\subsection{Dimensionless model and relevant dimensionless groups}

We scale the pressure ($\hat{p}$) and the deviatoric stress tensor ($\hat{\ubtau}$) with the buoyancy stress $\left( \hat{\rho}_f - \hat{\rho}_b \right) \hat{g} \hat{\ell}$, and the velocity vector ($\hat{\boldsymbol{u}}=(\hat{u},\hat{v})$) with the velocity $\hat{U}$:
\[
\left( \hat{\rho}_f - \hat{\rho}_b \right) \hat{g} ~\hat{\ell} = \hat{\mu} \frac{\hat{U}}{\hat{\ell}} \Rightarrow \hat{U} = \frac{\left( \hat{\rho}_f - \hat{\rho}_b \right) \hat{g} ~\hat{\ell}^2}{\hat{\mu}},
\]
which naturally arises from balancing the buoyancy stress with a viscous stress. In these equations, $\hat{\rho}_f$, $\hat{\rho}_b$, $\hat{g}$, $\boldsymbol{e}_g$ are respectively the densities of fluid, gas, the gravitational acceleration and its direction; $\hat{\mu}$ is the plastic viscosity. The length $\hat{\ell}$ is fixed by the dimensional bubble size. For our 2D planar flows, $\pi \hat{\ell}^2 = \text{meas} (X)$, and for our 3D axisymmetric flows, $(4/3) \pi\hat{\ell}^3 = \text{meas}(X)$, i.e.~considering the circle and sphere as reference geometries respectively. Thus, our scaled bubble domain $X$ will have area $\pi$, or volume $(4/3) \pi$.

The dimensionless Stokes and the constitutive equations are,
\begin{equation}\label{non-govern}
0 = - \boldsymbol{\nabla} p + \boldsymbol{\nabla} \boldsymbol{\cdot} \ubtau - \frac{1}{1-\rho} \boldsymbol{e}_g
\end{equation}
and
\begin{equation}\label{eq:const}
  \left\{
    \begin{array}{ll}
      \ubtau = \left( 1 + \displaystyle{\frac{Y}{\Vert\dot{\ubgamma}\Vert}} \right) \dot{\ubgamma} & \mbox{if}\quad \Vert \ubtau \Vert > Y, \\[2pt]
      \dot{\ubgamma} = 0 & \mbox{if}\quad \Vert \ubtau \Vert \leqslant Y.
  \end{array} \right.
\end{equation}
The 2 dimensionless groups are the density ratio $\rho = \hat{\rho}_b / \hat{\rho}_f$, and $Y = \hat{\tau}_y/(\Delta \hat{\rho} \hat{g} \hat{\ell})$, which is the yield number. Generally, we expect $\hat{\rho}_b \ll \hat{\rho}_f$, so that in practice $\Delta \hat{\rho} \approx \hat{\rho}_f$ and $\rho \approx 0$. In (\ref{eq:const}) $\dot{\ubgamma}$ is  the  rate of strain  tensor  and $\Vert \cdot \Vert$ is  the norm associated with the tensor inner product:
\[
\boldsymbol{c} \boldsymbol{:} \boldsymbol{d} =\frac{1}{2} \sum_{ij} c_{ij} ~d_{ij};
\]
e.g.~$\Vert \ubtau \Vert = \sqrt{\ubtau \boldsymbol{:} \ubtau}$.  The Cauchy stress tensor is $\ubsigma = -p \boldsymbol{I}+\ubtau$.

The velocity vector and tangential components of the traction (i.e.~$\left( \ubsigma \boldsymbol{\cdot} \boldsymbol{n} \right) \boldsymbol{\cdot} \boldsymbol{t}$) are continuous across the bubble surface $\partial X$.  Since the dynamic viscosity of the gas inside the bubble is generally negligible compared to the effective viscosity of yield-stress liquids, we assume:
\begin{equation}\label{eq:YL_tang}
\left( \ubsigma \boldsymbol{\cdot} \boldsymbol{n} \right) \boldsymbol{\cdot} \boldsymbol{t} \approx 0,~~~\mbox{on} ~\partial X.
\end{equation}
Thus, the tangential components of the traction vanish and the bubble can be regarded as inviscid: tangential velocities may slip and the normal velocity is continuous. The jump in the normal component of the traction is controlled by the dimensionless surface tension $\gamma~(= \hat{\gamma} / \hat{\rho}_f \hat{g} \hat{\ell}^2$; the inverse of the Bond number),
\begin{equation}\label{eq:YL_normal}
-p+p_b+(\ubtau \boldsymbol{\cdot} \boldsymbol{n}) \boldsymbol{\cdot} \boldsymbol{n} = \frac{\gamma}{\kappa}
\end{equation}
where
\[
\frac{1}{\kappa}=\frac{1}{\kappa_1}+\frac{1}{\kappa_2}
\]
and $\kappa_1$ and $\kappa_2$ are radii of curvature (in 2D $\kappa_2 = \infty$). In (\ref{eq:YL_normal}), $p_b$ is the pressure inside the bubble which can be considered as a constant in the inviscid \& incompressible limit.

The drag force on the bubble surface should balance the buoyancy of the bubble, $F$. In our 2D flows:

\begin{equation}
\hat{F} = \pi \hat{\ell}^2 \hat{\rho}_f \hat{g} = \int_{\partial X} \left[ (-\hat{p} \boldsymbol{1} + \hat{\ubtau}) \boldsymbol{\cdot} \boldsymbol{n} \right] \boldsymbol{\cdot} \boldsymbol{e}_g ~\text{d}\hat{S} = \int_{\partial X} \left( \hat{\ubsigma} \boldsymbol{\cdot} \boldsymbol{n} \right) \boldsymbol{\cdot} \boldsymbol{e}_g ~\text{d}\hat{S},
\end{equation}
or in dimensionless form:
\begin{equation}
F = \frac{\pi}{1 - \rho} = \int_{\partial X} \left( \ubsigma \boldsymbol{\cdot} \boldsymbol{n} \right) \boldsymbol{\cdot} \boldsymbol{e}_y ~\text{d}S .
\end{equation}

We will use these expressions later in \S \ref{sec:slipline}.

\subsection{Limit of zero flow}

In the present study, we are mainly interested in the yield limit of bubble motion, i.e.~conditions under which the bubble is held static. From the variational principles derived by \cite{dubash2004}, we can conclude that,
\begin{equation}
Y \int_{{\Omega} \setminus \bar{X}} \Vert \dot{\ubgamma} \Vert ~\mbox{d}\Omega \leqslant - \int_{{\Omega} \setminus \bar{X}} \boldsymbol{u} \boldsymbol{\cdot} \boldsymbol{e}_g ~\mbox{d} \Omega - \int_{\p X} \frac{\gamma}{\kappa} (\boldsymbol{u} \boldsymbol{\cdot} \boldsymbol{n}) ~\mbox{d}S .
\end{equation}
From left to right, the integrals above represent the plastic dissipation, denoted $j(\boldsymbol{u})$, the work done by buoyancy and by surface tension, denoted $L(\boldsymbol{u})$ and $T(\boldsymbol{u})$, respectively. In terms of these functionals, for $\boldsymbol{u} \not = 0$, we have:
\begin{equation}
Y \leqslant - \frac{\displaystyle \int_{{\Omega} \setminus \bar{X}} \boldsymbol{u} \boldsymbol{\cdot} \boldsymbol{e}_g ~\text{d} \Omega}{\displaystyle \int_{{\Omega} \setminus \bar{X}} \Vert \dot{\ubgamma} \Vert ~\text{d} \Omega} - \frac{\displaystyle \int_{\partial X} \frac{\gamma}{\kappa} \left( \boldsymbol{u} \boldsymbol{\cdot} \boldsymbol{n} \right) ~\text{d}S}{\displaystyle \int_{{\Omega} \setminus \bar{X}} \Vert \dot{\ubgamma} \Vert ~\text{d} \Omega}  \equiv \frac{L(\boldsymbol{u})}{j(\boldsymbol{u})} + \frac{T(\boldsymbol{u})}{j(\boldsymbol{u})}.
\end{equation}
Hence for a given shape of bubble, the critical yield number $Y_c$ above which the bubble will not rise is:
\begin{equation}\label{eq:Yc}
Y_c ~\equiv \sup_{\boldsymbol{v} \in \boldsymbol{V},~\boldsymbol{v} \neq 0} \left\{ - \frac{\displaystyle \int_{{\Omega} \setminus \bar{X}} \boldsymbol{v} \boldsymbol{\cdot} \boldsymbol{e}_g ~\text{d} \Omega}{\displaystyle \int_{{\Omega} \setminus \bar{X}} \Vert \dot{\ubgamma} \left( \boldsymbol{v} \right) \Vert ~\text{d} \Omega} - \frac{\displaystyle \int_{\partial X} \frac{\gamma}{\kappa} \left( \boldsymbol{v} \boldsymbol{\cdot} \boldsymbol{n} \right) ~\text{d}S}{\displaystyle \int_{{\Omega} \setminus \bar{X}} \Vert \dot{\ubgamma} \left( \boldsymbol{v} \right) \Vert ~\text{d} \Omega} \right\}
\end{equation}
where $\boldsymbol{v}$ is any admissible velocity field from the set $\boldsymbol{V}$ of all admissible fields. The first term on the right hand side represents the effect of the buoyancy stress on yielding: the numerator is the flux due to the bubble. This first term also appears in similar problems involving solid particles \citep{putz2010creeping, chaparian2017yield}, although there the interfacial conditions are different and $\rho \not= 0$. The second term represents the effect of the surface tension.

Since $\boldsymbol{v}$ is an admissible velocity field, it is also divergence free and the net flux through the bubble surface is consequently zero:
\[
\int_{\partial X} (\boldsymbol{v} \boldsymbol{\cdot} \boldsymbol{n})~\text{d}S = 0.
\]
We can split $\boldsymbol{v}$ into two components: the mean bubble speed $V_b$ and a perturbation $\boldsymbol{v}^{\prime}$:
\[
\boldsymbol{v} = \frac{- \displaystyle \int_{\Omega \setminus \bar{X}} (\boldsymbol{v} \boldsymbol{\cdot} \boldsymbol{e}_g) ~\text{d}\Omega}{\pi} ~\boldsymbol{e}_g + \boldsymbol{v}^{\prime}= V_b ~\boldsymbol{e}_g + \boldsymbol{v}^{\prime}~~~\text{on}~\partial X.
\]
The surface flux can now be rewritten as,
\[
\int_{\partial X} (\boldsymbol{v} \boldsymbol{\cdot} \boldsymbol{n}) ~\text{d}S = V_b \int_{\p X} (\boldsymbol{n}  \boldsymbol{\cdot} \boldsymbol{e}_g)~\text{d}S ~+  \int_{\p X} (\boldsymbol{v}^{\prime} \boldsymbol{\cdot} \boldsymbol{n}) ~\text{d}S = 0.
\]
It is clear that $\displaystyle \int_{\p X} (\boldsymbol{n}  \boldsymbol{\cdot} \boldsymbol{e}_g)~\text{d}S$ is zero, and therefore:
\[
\int_{\partial X} (\boldsymbol{v}^{\prime} \boldsymbol{\cdot} \boldsymbol{n})~\text{d}S = 0.
\]
Turning to the second term in (\ref{eq:Yc}), since $\kappa$ is not a constant for a general bubble shape along $\p X$,
\[ \int_{\p X} \frac{\gamma}{\kappa} (\boldsymbol{v} \boldsymbol{\cdot} \boldsymbol{n}) ~\text{d}S \not= 0, \]
in general. The only case in which this term is definitely zero, is a sphere (or a circular bubble in 2D). Physically this means that in the case of a spherical (or circular) bubble only buoyancy controls the yielding of the bubble; surface tension plays no role. This is intuitive for these equilibrium shapes, i.e.~we expect the surface tension to deform the bubble towards its equilibrium shape. For a general bubble shape, $Y_c$ is clearly also a function of the surface tension. Later, to analyze this effect the yield-capillary number is defined as:
\begin{equation}\label{eq:CaV}
Ca_Y = Y/\gamma = \frac{\hat{\tau}_y \hat{\ell}}{\hat{\gamma}}.
\end{equation}

\section{Methodology and benchmarking}
\label{sec:methods}

In overview, we use an augmented Lagrangian method coupled with an adaptive finite element method \citep{roquet2003adaptive} implemented in FreeFem++ \citep{MR3043640} to solve for the flow at fixed shape, $\gamma$ and $Y$. To determine $Y_c$ we then iteratively increase $Y$ until the flow stops. This basic procedure has been validated extensively in  previous studies \citep{roustaei2016non, chaparian2017yield, chaparian2020yield, chaparian2021sliding}, including those with particles. The mesh refinement nicely captures the yield surfaces after a few refinements.

\subsection{Computational method}

As we are concerned with finding the critical flow/no-flow condition (i.e.~$Y_c$), which represents a limiting balance between buoyancy, surface tension and plastic dissipation functionals, i.e.~(\ref{eq:Yc}), we compute the flows by solving the variational formulation of the exact non-smooth Bingham model. The alternative viscosity regularization methods are simple and effective in many cases, but have 2 drawbacks. First, characteristic features such as the yield surface shape are very sensitive to the regularization used \citep{Burgos1999}. Second, large errors can result when computing flows that are close to the critical no-flow limit, e.g.~as shown recently by  \citet{AHMADI2021}. The issue is that as the regularization parameter is taken to its limiting value, the velocity field of the regularized model does converge to the exact Bingham velocity, but there is no guarantee that the stress field will converge \citep{frigaard2005usage}.

The variational formulation of the Bingham fluid dates back to \cite{prager1954slow} and was formalised by \cite{duvaut1976inequalities}. They established two inequalities for the Stokes flow of Bingham fluid: a minimization based on velocity field (primal variable in optimization literature) and a maximization based on the stress field (dual variable). The two functionals are
\begin{align}
H(\boldsymbol{v})   &= \frac{1}{2} \int_{{\Omega} \setminus \bar{X}}\Vert \dot{\ubgamma} \Vert^2\ d\Omega + Y j(\boldsymbol{v}) - L(\boldsymbol{v}) - T(\boldsymbol{v}), \label{eq:VelMin} \\
K(\ubtau) &= \frac{1}{2} \int_{{\Omega} \setminus \bar{X}} \max(\Vert\ubtau\Vert-Y,0)^2\ d\Omega. \label{eq:StressMax}
\end{align}
The first term in (\ref{eq:VelMin}) is half of the viscous dissipation, denoted $a(\boldsymbol{v},\boldsymbol{v})$. An admissible velocity field $\boldsymbol{v} \in \mathcal{V}$ is any divergence free velocity satisfying the boundary conditions. An admissible stress field $\ubtau$ is any stress field satisfying the momentum equations (\ref{govern}) including the stress boundary conditions. A pair $(\ustar,\Tstar)$ that solve the Stokes problem will result in $H(\ustar)=-K(\Tstar)$. Note that the velocity is unique, but not the stress. For any other pair of admissible velocity/stress fields there is a difference between these values that is referred to as the duality gap. As discussed in \citet{treskatis2018practical}, the duality gap is the most reliable measure to make sure we have computed accurate stress fields as well as velocity fields, particularly close to the yield point.

\citet{Glowinski1981} developed the augmented Lagrangian (AL) method for solving the velocity maximization (primal formulation). The AL has been successfully used to solve various flows e.g.\ in ducts \citep{saramito2001adaptive}, around objects \citep{roquet2003adaptive}, and for inertial flows \citep{roustaei2015residualb}. Advances in computational optimization have led to a re-exploration of these flows recently, using algorithms that exploit the duality. A variation of the FISTA (fast iterative shrinkage-thresholding algorithm) was applied to the stress maximization formulation by \citet{treskatis2016accelerated}, which achieves a higher order of convergence in terms of the duality gap. Here we have used both AL and FISTA algorithms for our flows. This provides a basis to compare and cross validate. We have used the duality gap as the criteria for measuring the convergence of the solutions and find very little difference in converged solutions with either algorithm; see \S \ref{sec:AppendixA}.

\subsection{Benchmark flow}

\begin{figure}
\centerline{\includegraphics[width=0.7\linewidth]{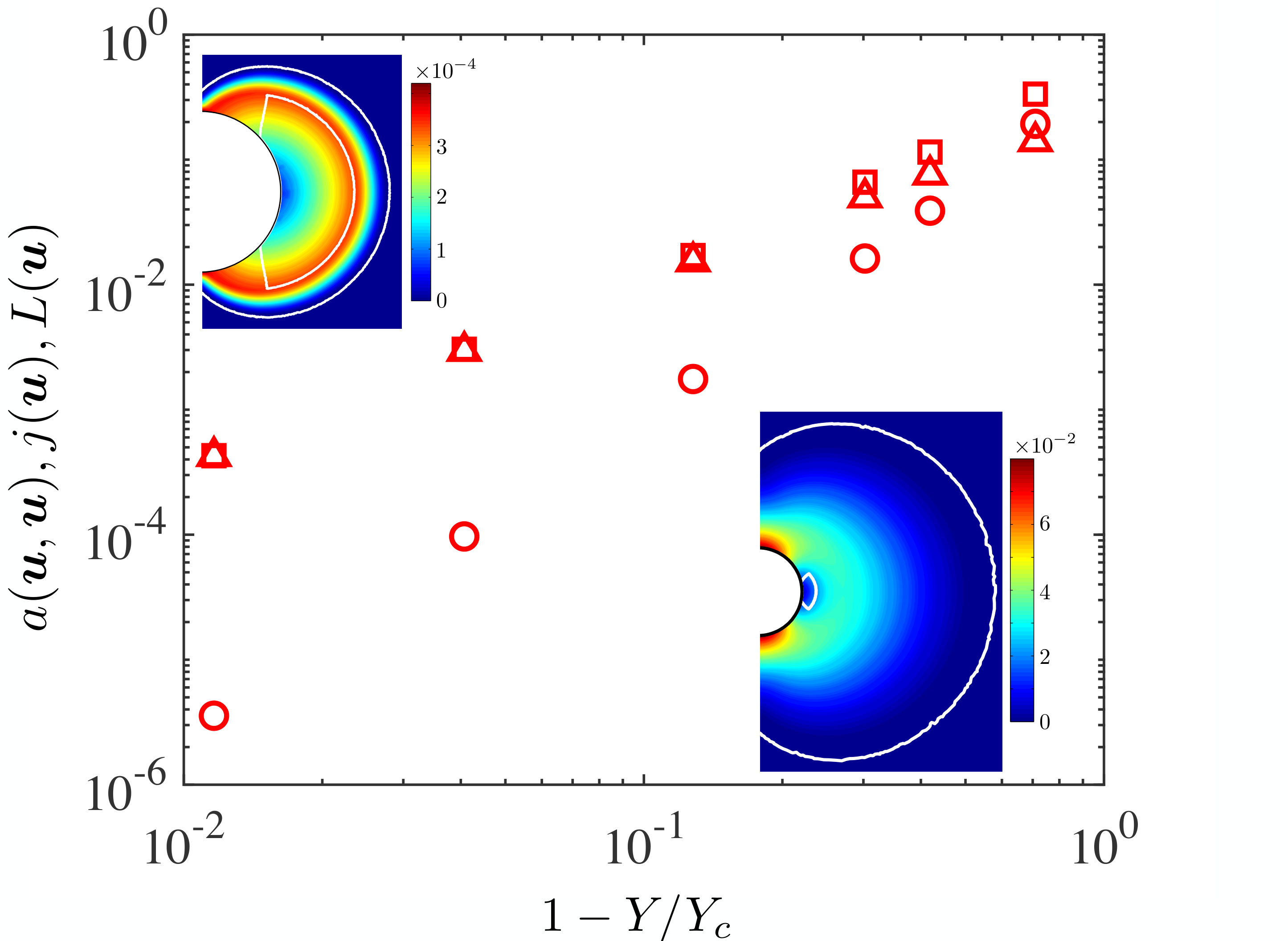}}
\caption{Viscous dissipation ($\bigcirc$), plastic dissipation ($\bigtriangleup$) and buoyancy work ($\square$) functionals for $\chi=1$ and $\gamma=0$. Insets illustrate the velocity field ($\vert \boldsymbol{u}  \vert$) for the cases of $1-Y/Y_c = 0.011$ (top left) and 0.419 (bottom right).}
\label{fig:functionals1}
\end{figure}

As a benchmark, we have computed 2D flows around circular and elliptical bubbles parameterised by an aspect ratio $\chi$, i.e.~the bubble surface is:
\begin{equation}\label{eq:ellipse}
  \chi x^2 + \frac{y^2}{\chi} = 1.
\end{equation}
The limiting solutions are explored fully later in the paper. Figure \ref{fig:functionals1} shows the viscous dissipation $a(\boldsymbol{u},\boldsymbol{u})$, plastic dissipation $j(\boldsymbol{u})$ and buoyancy work $L(\boldsymbol{u})$ functionals for $\chi=1$. All functionals decay with a power-law behavior as the yield number approaches the critical value. The viscous dissipation decays faster compared to the other functionals, in accordance with the computations of \citet{dimakopoulos2013steady}, and as must be true theoretically \citep{putz2010creeping}. Evidently, $ j(\boldsymbol{u}) \sim L(\boldsymbol{u})$ as they decay, as is implicit from (\ref{eq:Yc}) in the absence of surface tension. The velocity field for two cases of $1-Y/Y_c = 0.011,~~0.419$ are shown as insets. At yield numbers far away from the critical yield number (i.e. higher values of $1-Y/Y_c$), the yield surface surrounding the bubble is distant, and the yield surface at the bubble equator is small. As $Y$ gets closer to $Y_c$ the outer yield surface shrinks and the inner yield surface expands until they merge at $Y=Y_c$, where the bubble becomes entrapped.

\begin{figure}
\centerline{\includegraphics[width=0.7\linewidth]{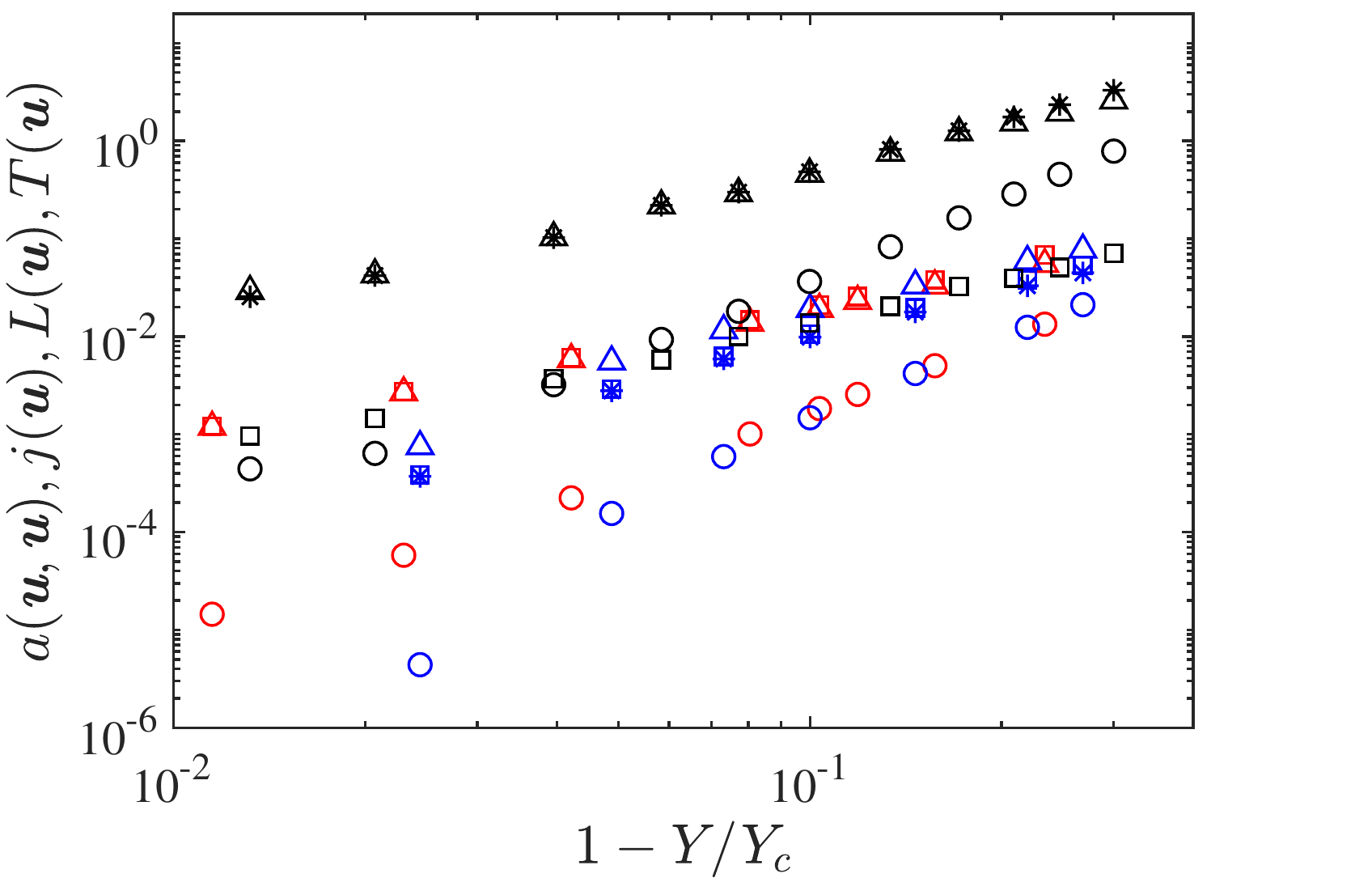}}
\caption{Viscous dissipation ($\bigcirc$), plastic dissipation ($\bigtriangleup$), buoyancy ($\square$) and surface tension work ($*$) functional for $\chi=2$. Cases of $\gamma=0$, $\gamma=1$ and $\gamma=10$ are shown in red, blue and black symbols respectively.}
\label{fig:functionals2}
\end{figure}

To explore the effects of surface tension we set $\chi=2$, so that the bubble is not in its equilibrium shape. Figure \ref{fig:functionals2} shows the surface tension functional $T(\boldsymbol{u})$ in addition to the previous functionals, for $\gamma=0, 1$ and 10. Again all functionals decay with a power-law behavior as the yield number approaches the critical value and the viscous dissipation decay is the fastest. In the absence of surface tension, buoyancy work mainly dissipates into plastic deformation. At $\gamma = 1$ we see an approximately equal $T(\boldsymbol{u}) \sim L(\boldsymbol{u})$, both balanced by $j(\boldsymbol{u})$. As the surface tension forces increase to dominate the flow (e.g.~$\gamma=10$), it can be seen that most of the plastic dissipation originates from balancing the surface tension work $T(\boldsymbol{u})$.

\subsection{Slipline analysis for 2D bubbles}\label{sec:slipline}

As the viscous dissipation is irrelevant to the yield limit, a natural direction to look for comparison is the theory of perfect plasticity, for which: $\Vert \hat{\ubtau}^* \Vert = \hat{\tau}_y$, since,
\begin{equation}\label{eq:PerfectPlastic}
\hat{\ubtau}^* = \frac{\hat{\tau}_y}{\Vert \hat{\dot{\ubgamma}}^*  \Vert} \hat{\dot{\ubgamma}}^* .
\end{equation}
Here we will use an asterisk to avoid mixing up with the viscoplastic variables. The unrestricted 2D perfectly-plastic flow can be transformed to an orthogonal curvilinear coordinate system ($\alpha$ and $\beta$) in which the directions coincide with the maximum shear stress ($\hat{\tau}_{\alpha \beta} = \pm \hat{\tau}_y$) directions \citep{hill1998mathematical,chakrabarty2012theory}. These orthogonal directions (i.e.~$\alpha-$ and $\beta-$directions) are the characteristic lines of the hyperbolic momentum equations in a 2D flow, known as the sliplines. Finding the sliplines does not require extensive computational effort and therefore this theory has been developed to compare with the viscoplastic `yield limit' for many different problems such as particle sedimentation \citep{chaparian2017cloaking,chaparian2017yield,chaparian2021sliding,hewitt2018slender}; swimming \citep{hewitt2017taylor,supekar2020translating}; and different studies in slump/dam-break type problems \citep{dubash2009final,liu2016two,chaparian2018lbox}. The solution procedure consists of postulating an admissible stress field by means of the sliplines and calculating the lower bound for the limiting force. An upper bound of the force also can be calculated using an admissible velocity field. The solution is more exact if the gap between the lower and upper bounds is small. This is the perfectly plastic version of the duality gap.

\cite{randolph1984limiting} proposed a slipline solution for flow around a laterally moving circular pile in soil to calculate the resistance. This solution has been used as a test problem in the context of 2D particle sedimentation in yield-stress fluids \citep[e.g.~see][]{chaparian2017yield}. One of the less explored parts of Randolph \& Houlsby's solution is the case of non-adhesive soil. In this case, the tangential shear stress at the pile surface is less than $\hat{\tau}_y$; namely $\left( \hat{\ubtau}^* \boldsymbol{\cdot} \boldsymbol{n} \right) \boldsymbol{t} = \alpha_s \hat{\tau}_y$ where $0 \leqslant \alpha_s \leqslant 1$ is the adhesion factor. This implies that one family of sliplines (say $\alpha-$family) makes an angle $\frac{1}{2} \left[ \frac{\pi}{2} - \sin^{-1} (\alpha_s) \right]$ with the pile interface. In other words, when $\alpha_s =1$, the $\alpha-$lines are tangential to the pile surface and when $\alpha_s=0$, $\alpha-$lines make $\frac{\pi}{4}$ angle with the tangent to the pile surface. In the viscoplastic fluid context, \citet{chaparian2021sliding} have shown that $\alpha_s$ represents the ratio of the `sliding yield-stress' to the fluid yield-stress. Hence, $\alpha_s = 1$ is indeed equivalent to imposing no-slip boundary condition on the object (e.g.~particle in a yield-stress fluid) and $\alpha_s=0$ is the same as having Navier slip (i.e.~zero sliding yield-stress) on the object.
It is natural therefore to apply the same solution to the bubble yield limit in the limit of $\gamma \to 0$ (as surface tension is not accounted for). A closer look at the boundary condition (\ref{eq:YL_tang}) reveals that for a 2D circular bubble, the solution is the same as \citet{randolph1984limiting} for $\alpha_s = 0$. We therefore briefly revisit the \citet{randolph1984limiting} solution for the case of 2D circular bubble and then extend it for an elliptical bubble with aspect ratio $\chi$.

\subsubsection{Circular bubble}

\begin{figure}
\centerline{\includegraphics[width=0.9\linewidth]{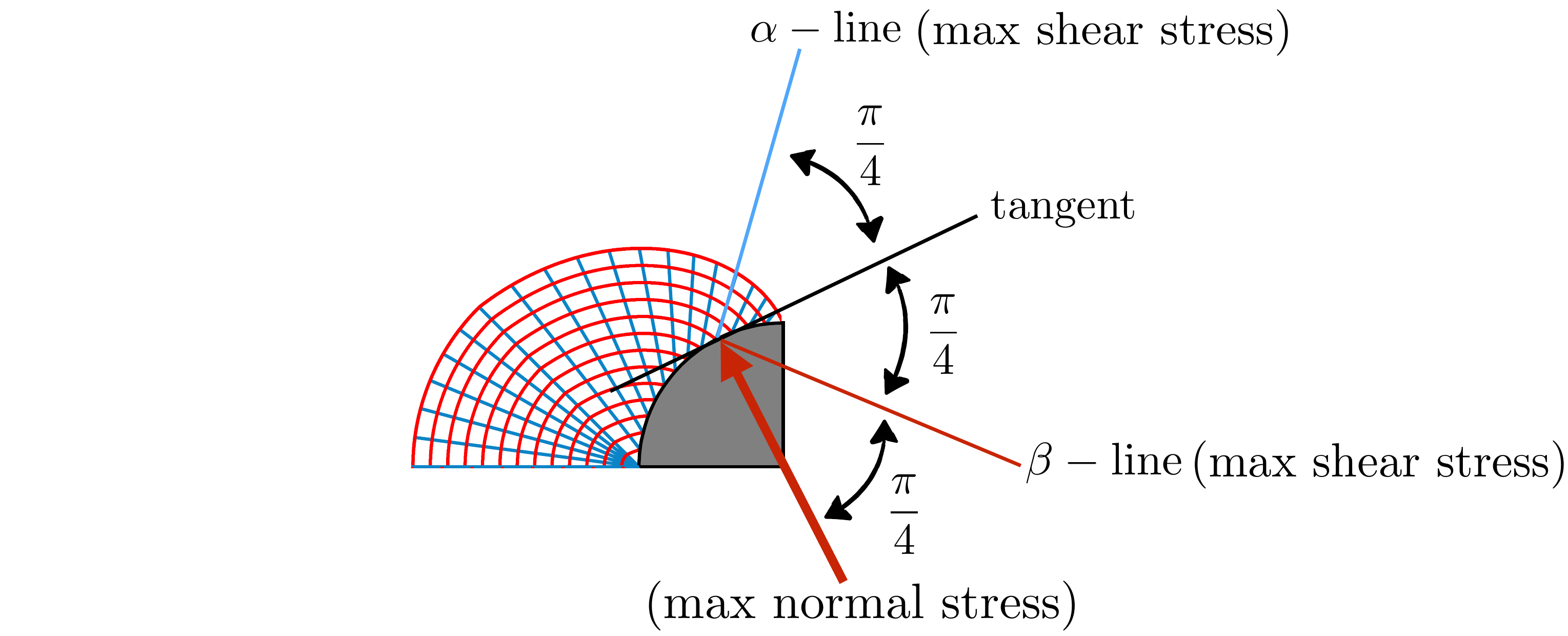}}
\caption{Schematic of the slipline network about a 2D circular bubble. A quarter of the domain is shown due to symmetry.}
\label{fig:schematic_slipline}
\end{figure}

This is covered in more detail in \citet{chaparian2021sliding}. The maximum normal stress direction is perpendicular to the bubble surface. This is directly dictated by the boundary condition on the bubble surface where the bubble pressure $p_b$, is constant along $\p X$. Hence, the $\beta-$lines make an angle $\pi / 4$ with the normal vector. As discussed above, the shear stress in the tangential direction is zero and therefore the $\alpha-$lines make an angle $\pi / 4$ with the tangent to the bubble surface; see figure \ref{fig:schematic_slipline}. From the symmetry constraints, along the vertical line of symmetry ($x=0$), the shear stress is zero ($\tau^*_{xy}=0$), while along the horizontal axis ($y=0$), it is maximum. Translating this into the slipline network, it means that the horizontal axis is itself an $\alpha-$line and all the sliplines that intercept the vertical axis should make an angle $\pi / 4$ with it.

We can construct an admissible stress field around the bubble with the aid of the sliplines and then integrate the force on the bubble surface in the direction $\boldsymbol{e}_g$ to find $[ \hat{F}^*_c ]_L$, the lower bound of the limiting force. The effect of buoyancy is absent in this admissible stress field, but by balancing the drag force with $\Delta \hat{\rho} \hat{g} \hat{A} = \Delta \hat{\rho} \hat{g} \pi \ell^2$, we can find the equivalent $Y_c$. The limiting plastic drag coefficient is defined as follows, where $\hat{\ell}_\bot$ is the linear length of bubble perpendicular to the flow direction, e.g.~twice the radius $\hat{R}$ for a circular bubble.
\begin{equation}
C_{d,c}^p = \frac{\hat{F}}{\hat{\tau}_y \hat{\ell}_\bot} = \frac{\hat{F}}{2 \hat{R}\hat{\tau}_y },
\end{equation}
and from the slipline analysis we find:
\begin{equation}
C_{d,c}^p = 6 + \pi~~\text{(for $\alpha_s=0$)}.
\end{equation}
Hence we balance buoyancy and the lower bound of the drag force,
\begin{equation}
\left[ \hat{F}^* \right]_L = 2 \hat{R} \hat{\tau}_y (6+\pi) = (\hat{\rho}_f - \hat{\rho}_b) \hat{g} \times \pi \hat{R}^2
\end{equation}
and find the upper bound of the critical yield number as:
\begin{equation}
\left[ Y_c \right]_U = \frac{\hat{\tau}_y}{\Delta \hat{\rho} \hat{g} \hat{R}} = \frac{\pi}{2 (6+\pi)} \approx 0.172 .
\end{equation}

%

\subsubsection{Ellipse}

For the planar elliptical bubble (\ref{eq:ellipse}), with aspect ratio $\chi$, we calculate the lower bound force from:
\begin{equation}\label{eq:lowerbound}
\left[ \hat{F}^* \right]_L = 4 \int_0^{\frac{\pi}{2}} \hat{\tau}_y \hat{\ell}_{\bot} \left( \bar{\sigma} + \frac{3\pi}{2} + 1 - 2 \zeta \right) \sqrt{\chi^{-1} \sin^2{\theta_1}+\chi \cos^2{\theta_1}} \cos{\zeta}~ \mbox{d} \theta_1
\end{equation}
where, $\theta_1 = \tan^{-1} \left[ \chi^{-1} \tan{\theta} \right]$ and $\zeta = \tan^{-1} \left( \chi \cot \theta_1 \right)$. Details of this type of calculation can be found in \cite{chaparian2017yield} and \cite{chaparian2021sliding}.

This force lower bound converts to:
\begin{equation}\label{eq:upperbound}
\left[ Y_c \right]_U = \frac{\pi}{4 \left[ \hat{F}^* \right]_L}
\end{equation}

\subsubsection{Comparing slipline and viscoplastic solutions}

\begin{figure}
\centerline{\includegraphics[width=0.7\linewidth]{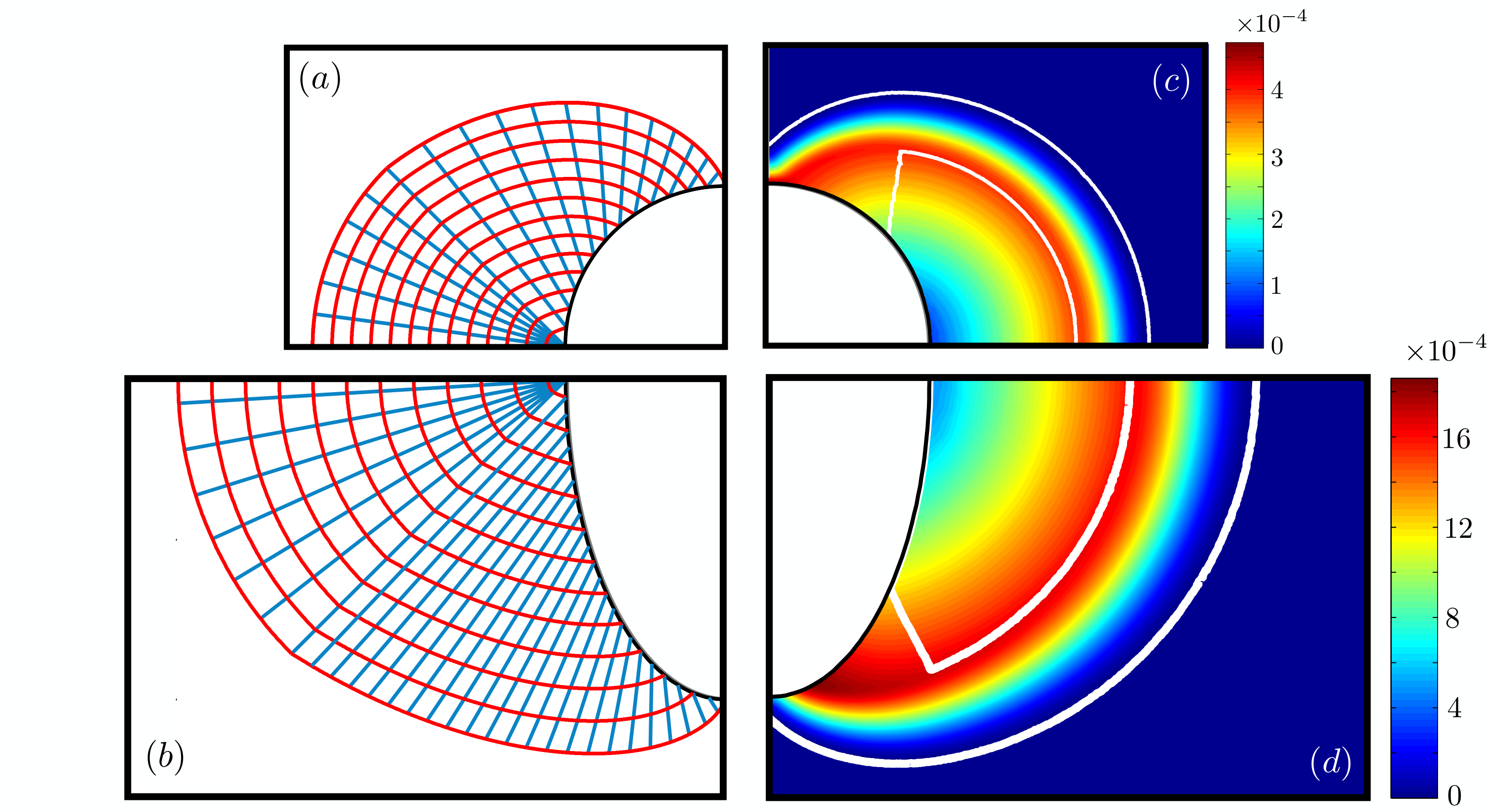}}
\caption{(a,b) Slipline network about a 2D circular bubble and a 2D ellipse with $\chi=2$. (c,d) Velocity $\vert \boldsymbol{u} \vert$ contour; $Y=0.17$ and $Y=0.255$ for the circular and elliptical bubble, respectively. Unyielded regions are marked with white solid lines.}
\label{fig:sliplinecompare}
\end{figure}

Figure \ref{fig:sliplinecompare} compares the slipline network with the viscoplastic solution for $\chi = 1,~2$, for $Y$ just below $Y_c$. Although not precisely the same, the envelope of the sliplines approximates the yielded region of the viscoplastic fluid. The white line denotes the yield surface in the viscoplastic fluid and we see that the fluid is unyielded over a large part of the bubble surface, extending from the horizontal axis.

\begin{figure}
\centerline{\includegraphics[width=0.5\linewidth]{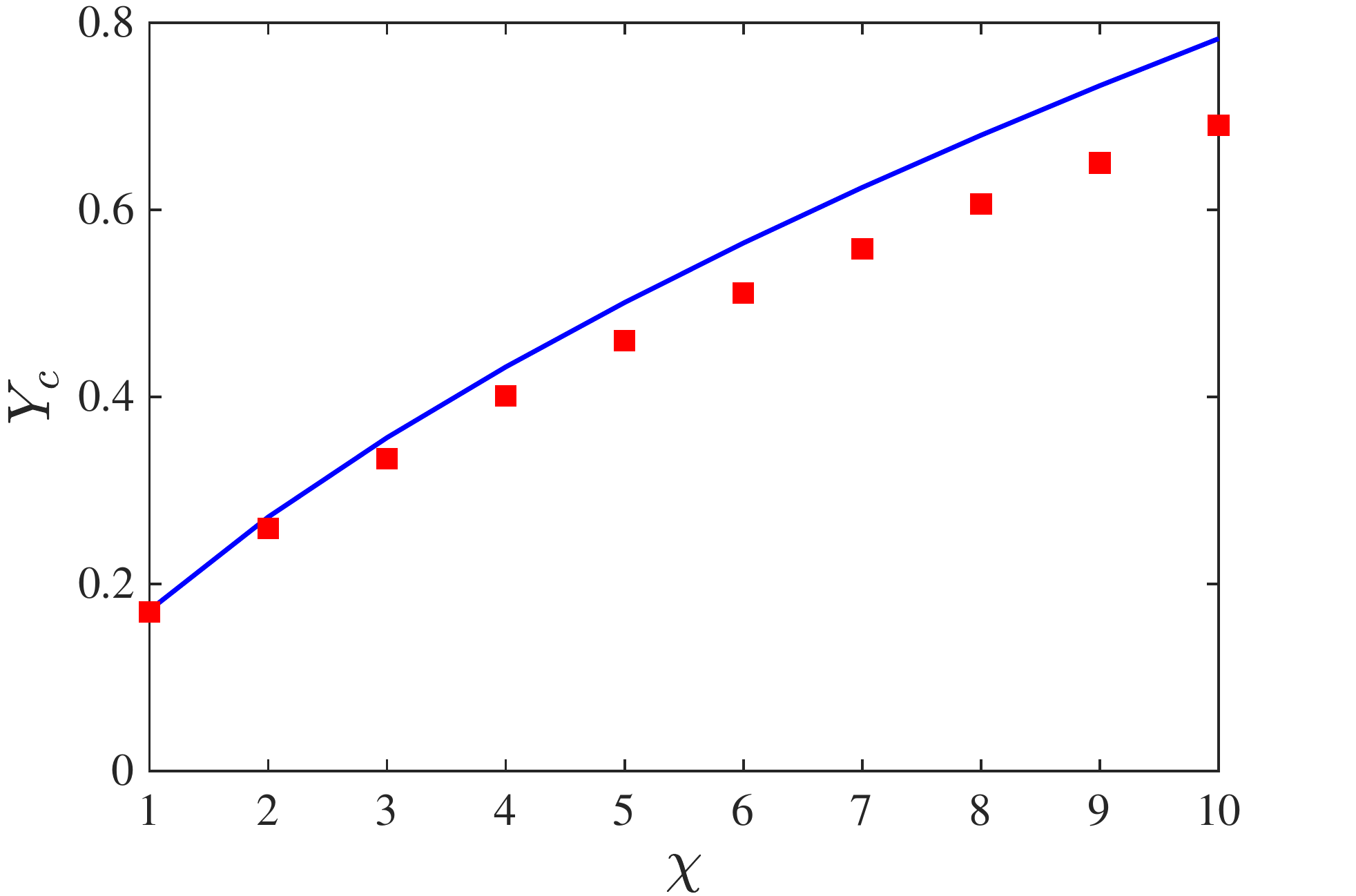}}
\caption{$Y_c$ versus the aspect ratio for $\chi \geqslant 1$; the symbols represent the viscoplastic computations and the line is expressions (\ref{eq:lowerbound}) \& (\ref{eq:upperbound}).}
\label{fig:Yccompare}
\end{figure}

As we increase $Y$ we find $Y_c$ computationally for the viscoplastic flow, which can be compared with that from  (\ref{eq:lowerbound}) \& (\ref{eq:upperbound}). We find: $\chi=1:~Y_c \approx 0.172$, $\chi=2:~Y_c \approx 0.27$ and  $\chi=5:~ Y_c \approx 0.5$ from the slipline method, compared with: $\chi=1:~Y_c \approx 0.172$,  $\chi=2:~Y_c \approx 0.267$ and $\chi=5: ~Y_c \approx 0.460$, from the limiting viscoplastic flow computations. Figure \ref{fig:Yccompare} shows this comparison graphically over a wider range of $\chi$, from which we can see a growing discrepancy between the two limits as $\chi$ increases.

There are two reasons for this discrepancy. First, we should note that the slipline solutions and limiting viscoplastic solutions are not quite simply the same. Although both flows limit to zero motion at $Y_c$, the stress fields are not the same. In the viscoplastic solution the yield surfaces bound regions within which the stress is at or below the yield-stress. In the perfectly plastic solution the stress is constrained to be at the yield value everywhere within the slipline envelope. The slipline stress field can be used to construct an admissible stress field for the viscoplastic problem (if it can be extended outside the slipline envelope), and using the stress maximization principle one can infer that this will lead to an upper bound for the viscoplastic $Y_c$, as is observed.

The second reason for the discrepancy is that although widely believed that the \citet{randolph1984limiting} solution is exact (i.e.~the lower and upper bounds of the drag are the same for the whole range of  $\alpha_s$), this is not the case. An issue was first detected by \citet{murff1989pipe}: for $\alpha_s<1$, there is a region in the vicinity of the pile surface in which the rate of strain is negative and its absolute value should consequently be taken into account in calculating the upper bound, to avoid negative plastic dissipation. If one does so, then there is a discrepancy between the lower and upper bounds, i.e.~for $\alpha_s<1$ the \citet{randolph1984limiting} solution is not exact for the perfectly plastic problem. \citet{martin2006upper} postulated two new velocity fields which improved the  upper  bound  predictions to some extent. See \citet{chaparian2021sliding} for a detailed comparison with the viscoplastic flow about a solid particle. Also \citet{supekar2020translating} have shown that the \citet{martin2006upper} solution is indeed superior compared to the lower bound solution.

Note that this does not affect the results here in figure \ref{fig:Yccompare}, which are taken from (\ref{eq:lowerbound}) \& (\ref{eq:upperbound}) and use the slipline stress field (the lower bound). Indeed, our conclusion is that the slipline method is a valuable check on computed limiting solutions. It exhibits very similar geometric features and often gives a close estimate to $Y_c$. A similar conclusion was reached in our study of solid particles \citep{chaparian2017yield}, for which the \citet{randolph1984limiting} slipline solution is both correct and exact ($\alpha_s=1$).

\section{Results for 2D bubbles}
\label{sec:Results}

\subsection{Elliptic bubbles}

Here we investigate the behavior of critical yield number ($Y_c$) with respect to both aspect ratio  ($\chi$) and the dimensionless surface tension ($\gamma$) for 2D elliptical bubbles described by (\ref{eq:ellipse}). The range of aspect ratios considered are between 0.1 -- 10 and the range of surface tensions considered are between 0 -- 10. Regarding realistic values of ($\chi$) and ($\gamma$), there are two main sources of information. First we refer to the discussion of different static bubble shapes in \S \ref{sec:intro}. Second, we can look at experimental studies of rising bubbles in yield-stress fluids \citep{dubash2007propagation, sikorski2009motion, lopez2018rising, pourzahedi2021eliminating}, where they find aspect ratios, 0.2 -- 5, and $\gamma$ in the range 0.01 -- 0.5. Evidently, experimental bubble shapes are not elliptical (nor planar) and we also note that moving bubbles in experiments tend to be larger than stationary (increased buoyancy). Thus, our range of parameters is practically motivated. In order to find $Y_c$, we vary the yield number until the flow field reaches zero everywhere in the domain using the computational methods discussed in previous section. 

\begin{figure}
\centerline{\includegraphics[width=0.7\linewidth]{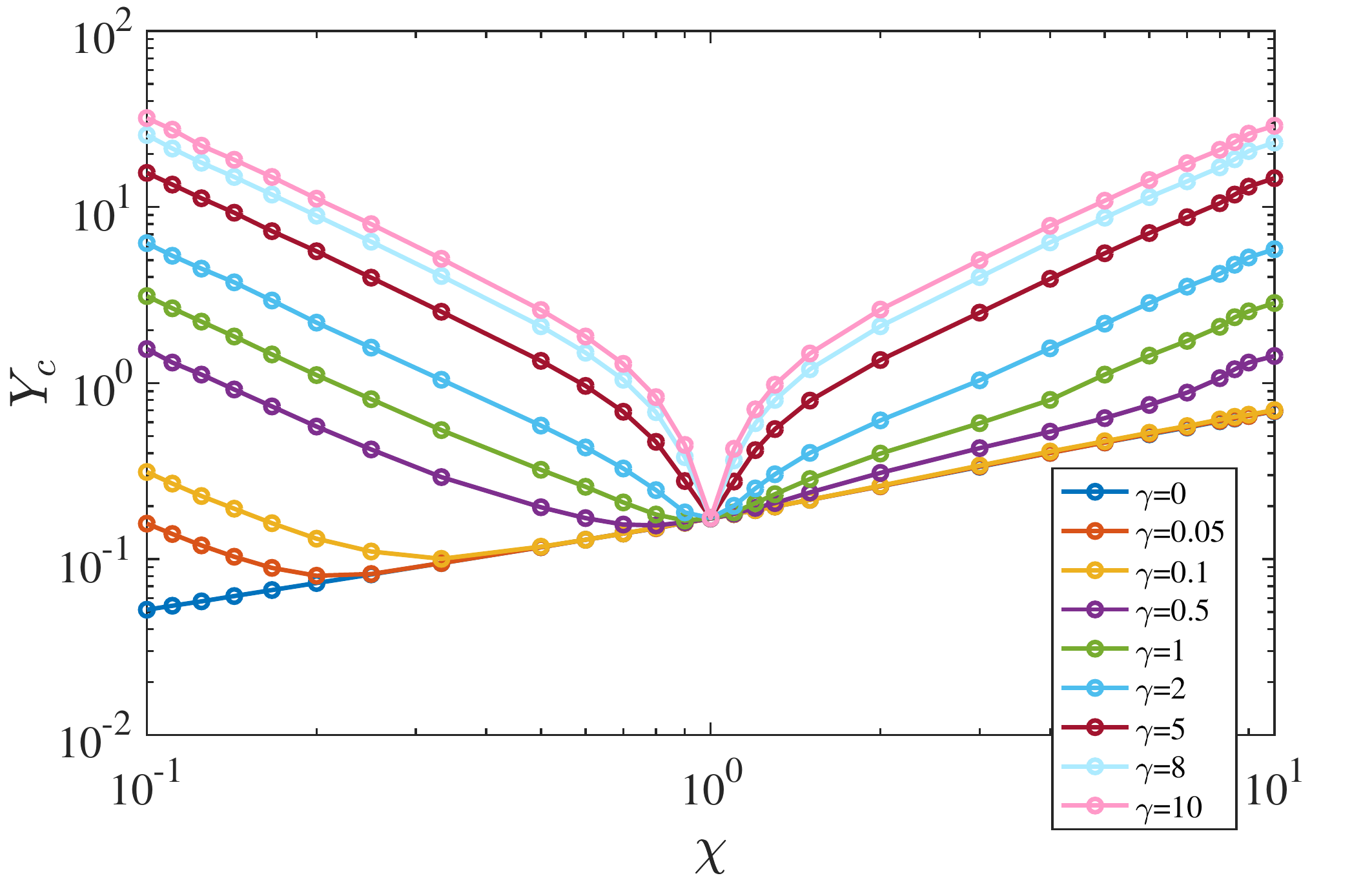}}
\caption{Critical yield number against aspect ratio for various surface tensions for elliptic bubbles.}
\label{fig:Y-Chi}
\end{figure}

Figure \ref{fig:Y-Chi} shows how the critical yield number varies with aspect ratio for different surface tension parameter $\gamma$. For $\gamma = 0$, $Y_c$ increases with $\chi$: prolate bubbles  ($\chi > 1$) can flow easier compared to oblate bubbles ($\chi < 1$). One physically intuitive interpretation of the increases is that for increasing $\chi$ the bubbles create a static pressure differential proportional to their height: a higher yield-stress is consequently required to overcome the stresses created. The height of the bubble scales with $\sqrt{\chi}$, which is approximately the slope of the $\gamma = 0$ curve. Although attractive in its simplicity, the same explanation does not account for our axisymmetric results later, which suggests that there is more going on here and a different explanation needed. We consider $\chi < 1$ and $\chi > 1$ separately.

For $\chi < 1$, it is apparent that in order for the bubble to propagate the fluid must move around an increasingly wide obstacle as $\chi \to 0$. Primarily the bubble \emph{pushes} the fluid out of the way, suggesting that normal stresses at the interface are important and tangential stresses less relevant. In this sense oblate bubbles and particles may be similar in their yield limit behaviour. In \citet{chaparian2017yield}, the yield limit of planar elliptical particles for $\chi \ll 1$ varies as $Y_c \sim \chi^{0.5}$, just as here. However, the rigid particle displaces a triangular region of unyielded fluid ahead of it, which then pushes a ring of fluid around the particle at constant speed; see figure 10c in \citet{chaparian2017yield}. This simple flow structure allows us to estimate $Y_c$ directly for the particle.

When we explore the flow around the bubble as $Y \sim Y_c$, for $\chi < 1$, we observe that the yielded fluid is confined within 2 disks of approximate radius $\sqrt{2/\chi}$, centred at the ends of the bubble. This feature is just like the flow around elliptical particles. Unlike the rigid particle, the flow is significantly less structured, without sharp gradients. This prevents us from easily deriving an approximation to $Y_c$, but the order of $Y_c$ as $\chi \to 0$ can be estimated. From equation (\ref{eq:Yc}), we see that $Y_c \approx L(\boldsymbol{u})/j(\boldsymbol{u})$ for the limiting solutions $\boldsymbol{u}$. If $U_b$ represents the mean bubble rise velocity, then we also have $L(\boldsymbol{u}) \sim U_b $, due to incompressibility. The bubble width is $2a$, also giving the length-scale of the yielded region. The strain rate is likely to scale with $U_b /a$ within the yielded region, leading to $j(\boldsymbol{u}) \sim U_b/a \times a^2 = a U_b$. Dividing through we have: $Y_c \approx L(\boldsymbol{u})/j(\boldsymbol{u}) \sim 1/a \sim \chi^{0.5}$, as we have observed.

In contrast, for $\chi > 1$ we cannot expect that the bubble yield limit behaves as the particle yield limit. The particle is strongly influenced by a thin yielded boundary layer \citep{chaparian2017yield,hewitt2018slender}. For $Y \sim Y_c$, we find that for tall bubbles much of the bubble surface is covered by an unyielded plug that moves slowly downwards as the bubble slips against it. This plug region is surrounded by a yielded shear layer and then bounded by the outer yield envelope. In the shear layer the shear stress $\Vert \ubtau \Vert$ exceeds the yield-stress $Y$. We may suppose that $\Vert\dot{\ubgamma}\Vert  \sim (\Vert \ubtau \Vert - Y) \propto d K$, where $d$ represents the distance from the outer yielded envelope to the inner plug, and $K$ is the size of the pressure gradient oriented along the flow in the shear layer. Thus, now $j(\boldsymbol{u}) \sim d^2 K \chi^{0.5}$, if we estimate the shear layer as having size $d \times \chi^{0.5}$. For the velocity within the yielded envelope, we integrate once more to find $|\boldsymbol{u}| \propto d^2 K$ and now integrate over the entire region within the yielded envelope to give: $L(\boldsymbol{u}) \sim d^2 K \chi$, i.e. not only across the width of the shear layer. Note that both the height and width of the yielded envelope appear to scale with $\chi^{0.5}$. Thus again we find $Y_c \approx L(\boldsymbol{u})/j(\boldsymbol{u}) \sim \chi^{0.5}$.

\begin{figure}
\centerline{\includegraphics[width=0.7\linewidth]{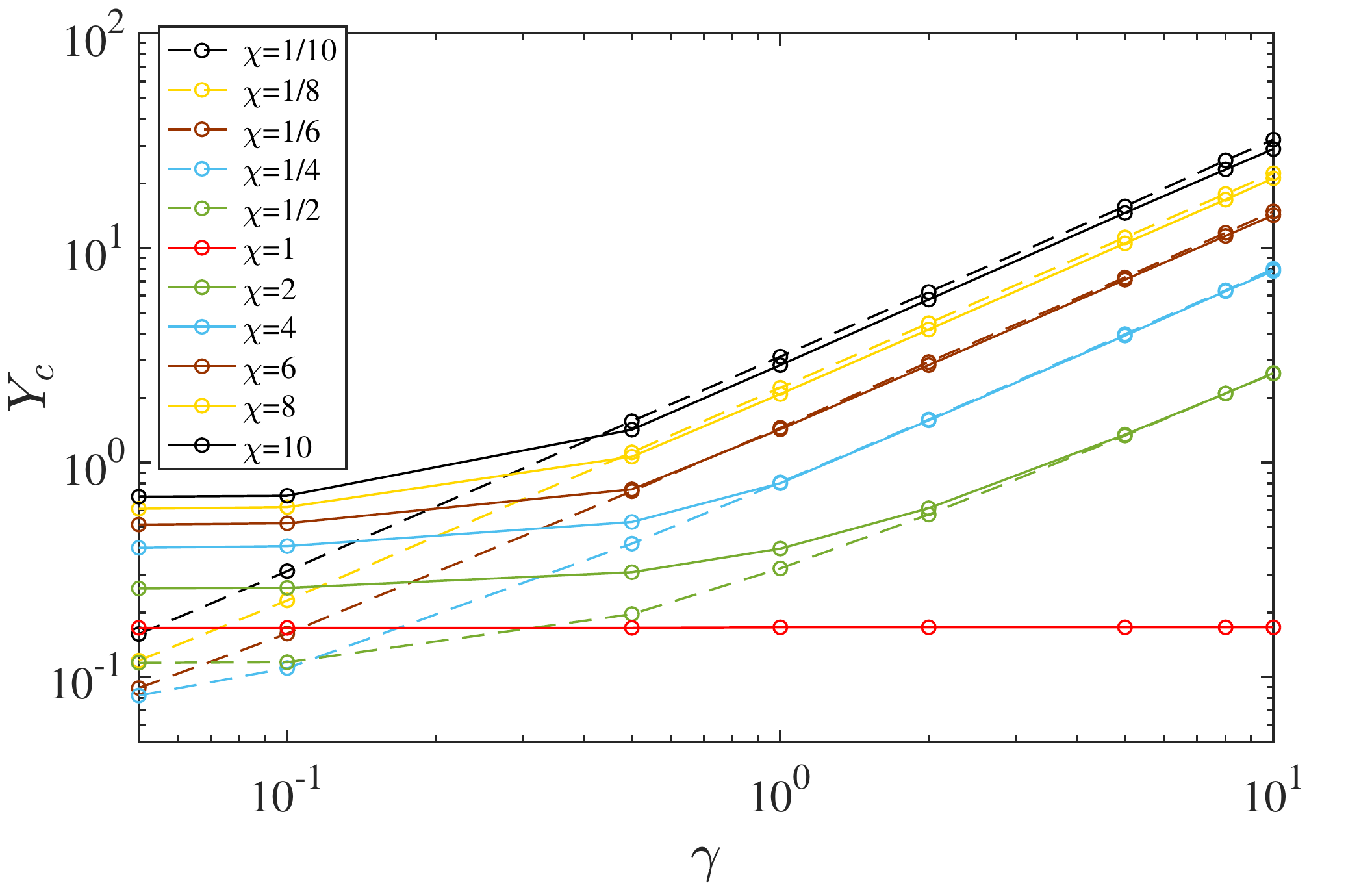}}
\caption{Critical yield number against surface tension for various aspect ratios of elliptic bubbles. $\chi<1$ is plotted in dotted lines.}
\label{fig:Y-gamma}
\end{figure}

\begin{figure}
\centerline{\includegraphics[width=0.7\linewidth]{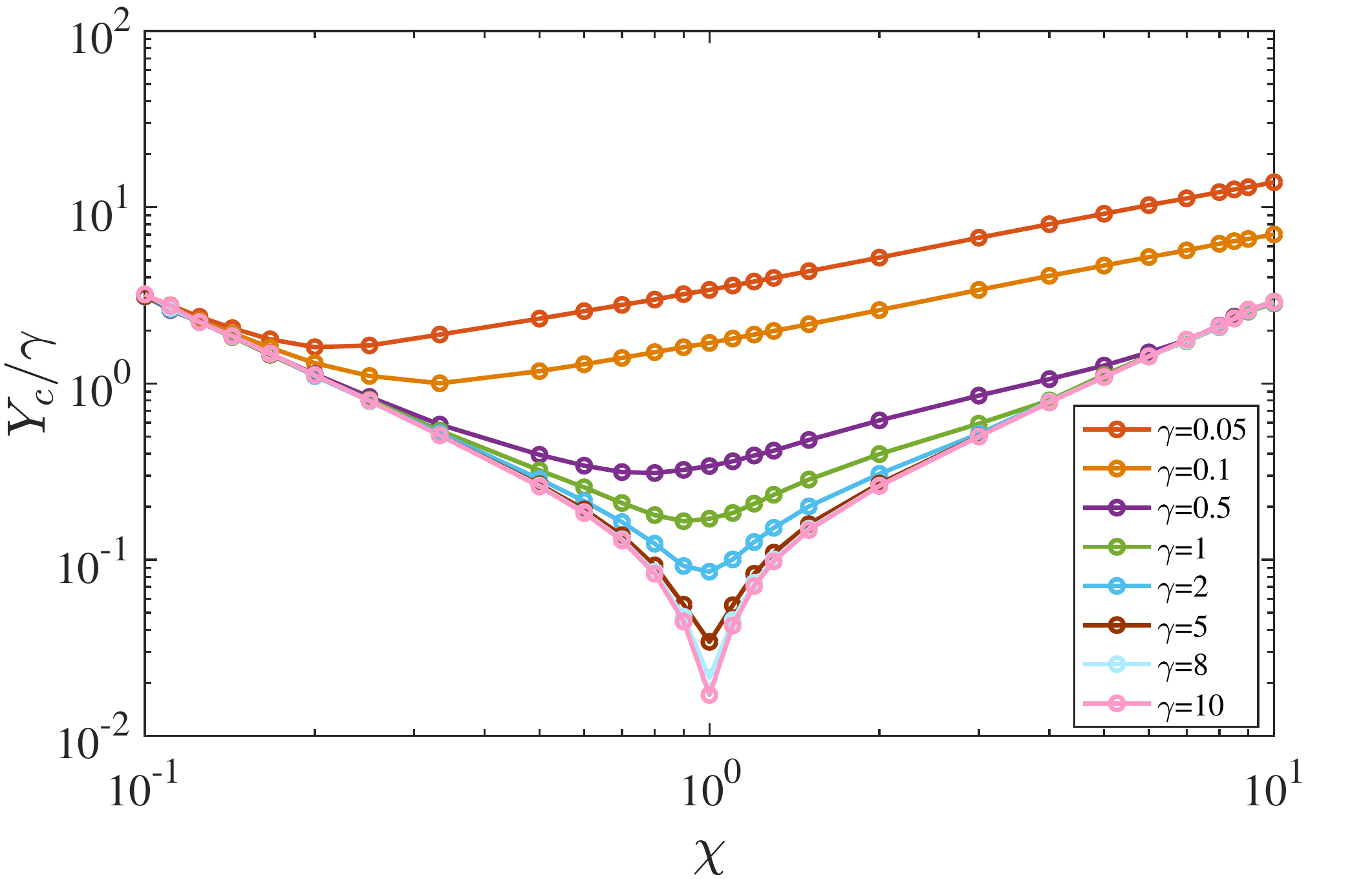}}
\caption{Yield-capillary number against aspect ratio for various surface tensions for elliptic bubbles.}
\label{fig:Y_gamma-Chi}
\end{figure}

For $\gamma > 0$ we see a departure from the $\gamma = 0$ curve. Surface tension has no effect at $\chi = 1$, and deviation from the $\gamma = 0$ curve is more abrupt for larger $\gamma$.
As $\gamma$ increases, the curves become increasingly symmetrical about $\chi=1$, meaning that deformation occurs mainly due to surface tension. This behavior is better illustrated in figure \ref{fig:Y-gamma} where $Y_c$  is plotted with respect to $\gamma$ for various aspect ratios. It can be observed that the $Y_c$ values for specific $\chi$ and $1/\chi$ collapse on each other for high values of $\gamma$ (e.g.~$\chi=2, 0.5$).
The overlap of the curves for $\chi$ and $1/\chi$ in figure \ref{fig:Y-gamma} suggests that the high curvature of the more pointed ends of the bubble leads to a dominant effect on the stress field that must be resisted by the yield-stress. Therefore, eventually we expect to find a direct balance between yield-stress and surface tension.
The ratio of yield number to surface tension is the yield-capillary number discussed earlier. The critical value with respect to aspect ratio is plotted for various values of $\gamma$ in figure \ref{fig:Y_gamma-Chi}.
As $\gamma\to\infty$ the curves appear to collapse to a limiting curve, confirming this interpretation. We may also evaluate the maximal curvature of the bubble, as a function of $\chi$, which occurs at the top and bottom for $\chi>1$ and at the sides for $\chi < 1$. We find that the maximal contribution of the surface tension to the jump in pressures is given by $\gamma \chi^{1.5}$ for $\chi>1$, and by $\gamma \chi^{-1.5}$ for $\chi<1$. The symmetry for large/small $\chi$ is evident in the planar bubble shape. On returning to figure \ref{fig:Y-Chi}, we note that for large $\gamma$ the slopes of the $Y_c$-curves align well with $\sim \chi^{1.5}$ for $\chi \gg 1$, and $\sim \chi^{-1.5}$ for $\chi \ll 1$.

The effect of the surface tension on the velocity field around moving bubbles, with $\chi=2$ and $\chi=0.5$, is depicted in figure \ref{fig:Chi=2,0.5}a--h.  In all cases, the yield number is selected to satisfy $1-Y/Y_c =0.1$. It can be seen that velocity magnitudes increase with surface tension for both aspect ratios, suggesting that larger yield number is required to immobilize the bubble at higher surface tension values. We emphasize that these velocity fields do not represent steadily moving bubbles, but are simply Stokes flow solutions for these specific shapes. In the case of $\gamma=0$ for both aspect ratios the flow field, as well as the unyielded regions are symmetrical in the upper and lower halves of the bubble (figures \ref{fig:Chi=2,0.5}a \& e). For the case of $\chi=2$, in addition to the outer field unyielded region, an unyielded region forms around the bubble equator at low surface tensions, and the yielded fluid circulates from the top to bottom in the pathway between the two regions (figure \ref{fig:Chi=2,0.5}b). Interestingly the unyielded region attached to the equator shrinks as surface tension increases, and the outer unyielded region surrounds the lower tip (figure \ref{fig:Chi=2,0.5}c). This can be explained by considering the role of surface tension in defining the bubble shape. Surface tension tends to push the bubble boundary into its equilibrium shape (i.e. circle). In the lower half, buoyancy and surface tension forces add up as they both push upwards. However in the upper half of the bubble, the surface tension force changes direction and pulls downwards.  Therefore, they cancel each other out partly. Thus, higher yield-stress is required to immobilize the lower half of the bubble compared to the upper half. Further increasing the surface tension creates stronger forces at the tips, which again starts to yield the fluid around the upper tip of the bubble (figure \ref{fig:Chi=2,0.5}d). The unyielded region in the lower half is larger than the upper half due to the same reason explained above.

For the case of $\chi=0.5$, similarly, the unyielded region attached to the bubble equator shrinks as surface tension increases. The shape of the unyielded region can be justified using the same logic as before. In the upper half both the surface tension and buoyancy forces are pointing upwards. However in the lower half buoyancy is still pointing upwards while the surface tension is acting downwards, therefore they counteract each other. Thus the upper half requires more yield-stress to become immobilized (figure \ref{fig:Chi=2,0.5}f). Further increasing the surface tension yields the fluid around the bottom half and induces a new unyielded region.

\begin{figure}
\centerline{\includegraphics[width=\linewidth]{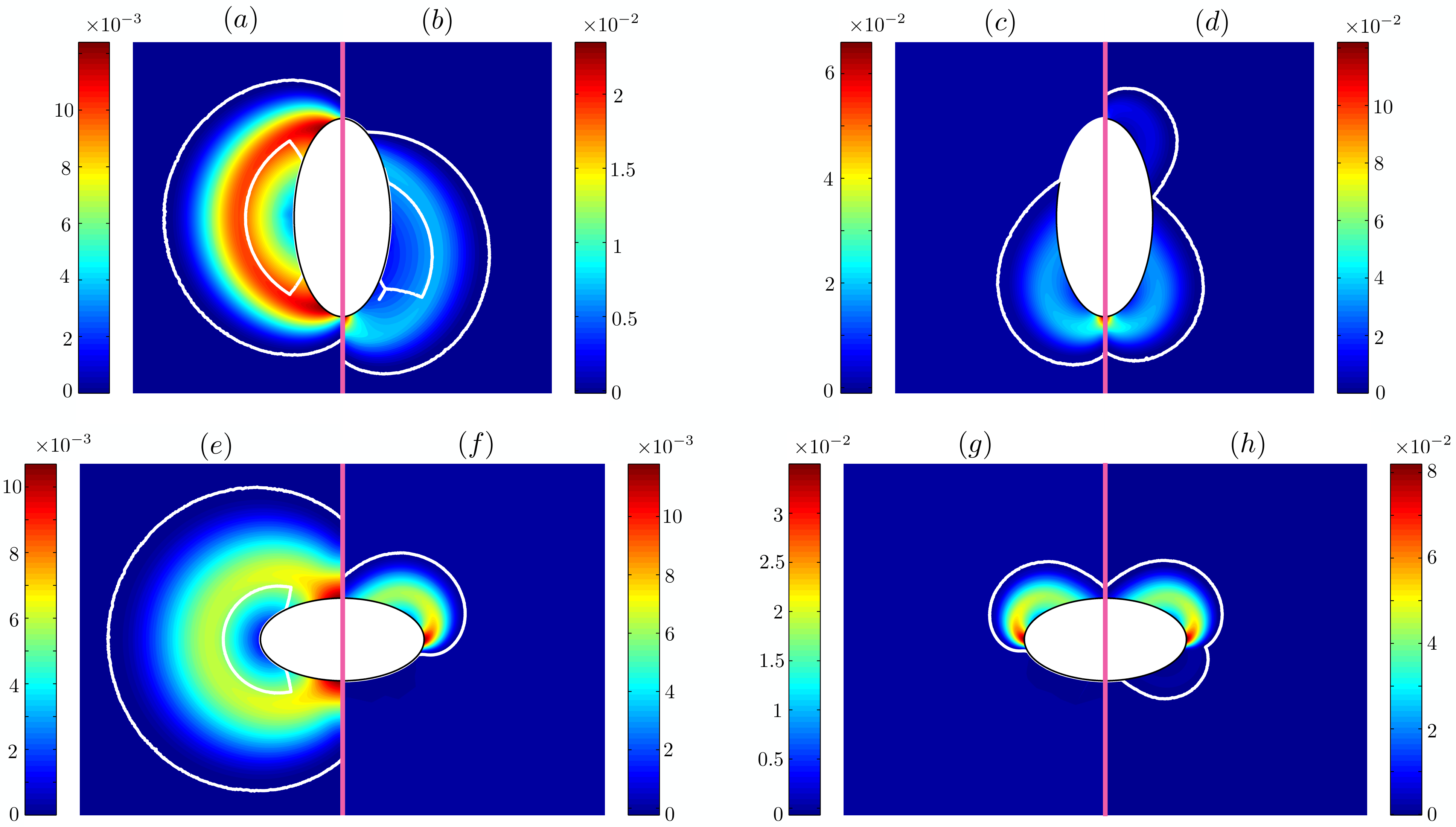}}
\caption{Velocity field around an elliptical bubble for the cases of $\chi=2$ (top row) and $\chi=0.5$ (bottom row), for $\gamma$ equal to: 0 (a, e), 1 (b, f),  5 (c, g) and 10 (d, h). In all cases, the yield number satisfies $1-Y/Y_c =0.1$.}
\label{fig:Chi=2,0.5}
\end{figure}

\subsection{Long bubbles}

\begin{figure}
\centerline{\includegraphics[width=0.6\linewidth]{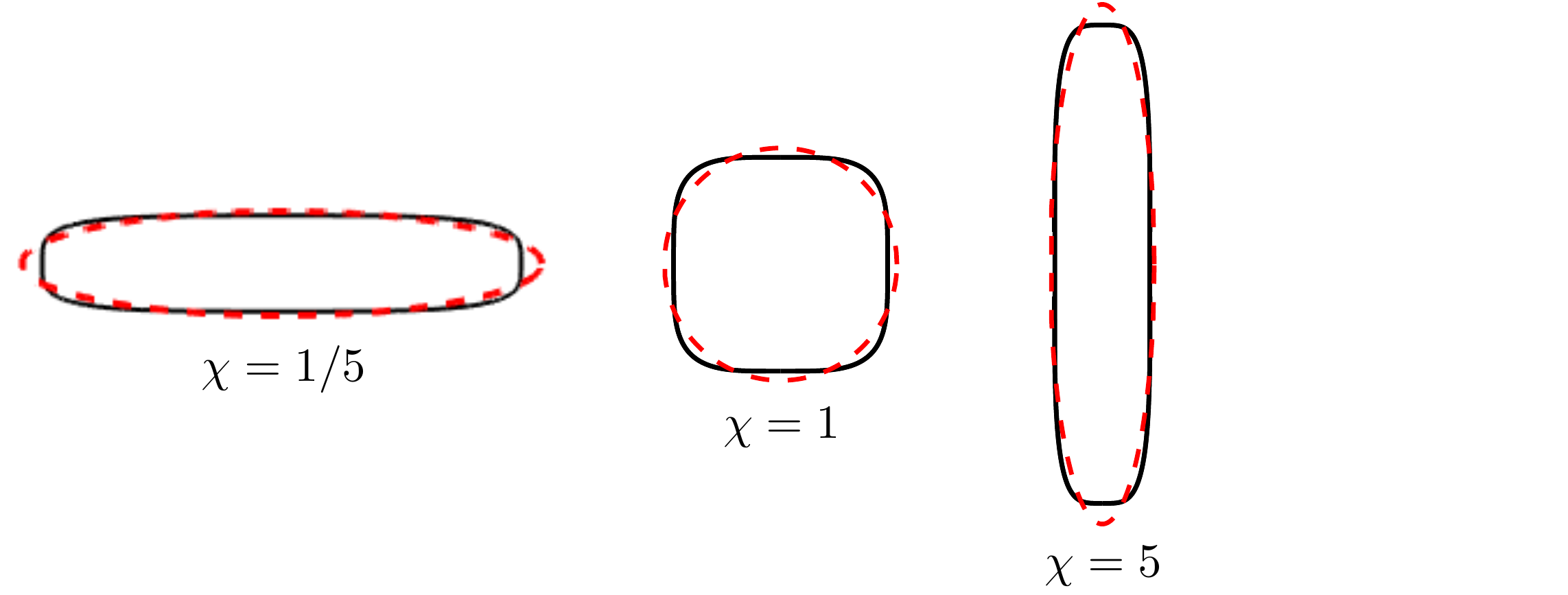}}
\caption{Illustration of quartic (black line) and elliptical bubble (discontinuous red) shapes at various aspect ratios.}
\label{fig:schematic_quartic}
\end{figure}

To give a second geometry we explore bubbles with quartic profile. To clarify, the interface is given by: \begin{equation}\label{eq:quartic}
  (x/a)^4+(y/b)^4=1,
\end{equation}
with an area equal to $\pi$. We denote the aspect ratio $b/a = \chi$ and again investigate the variation of $Y_c$, with $\chi$ and $\gamma$. A selection of shapes are shown in figure \ref{fig:schematic_quartic}. Here we just note that $\chi=1$ does not give a circular shape and in general these bubbles are more \emph{rectangular}, with rounded corners, in comparison to the elliptical bubbles. As an aside, we first tried some computations of long straight-sided bubbles with semi-circular ends. However, the jump in curvature (between circular cap and straight sides) created unphysical behavior when $\gamma > 0$, whereas the quartic profile is smooth.

\begin{figure}
\centerline{\includegraphics[width=0.7\linewidth]{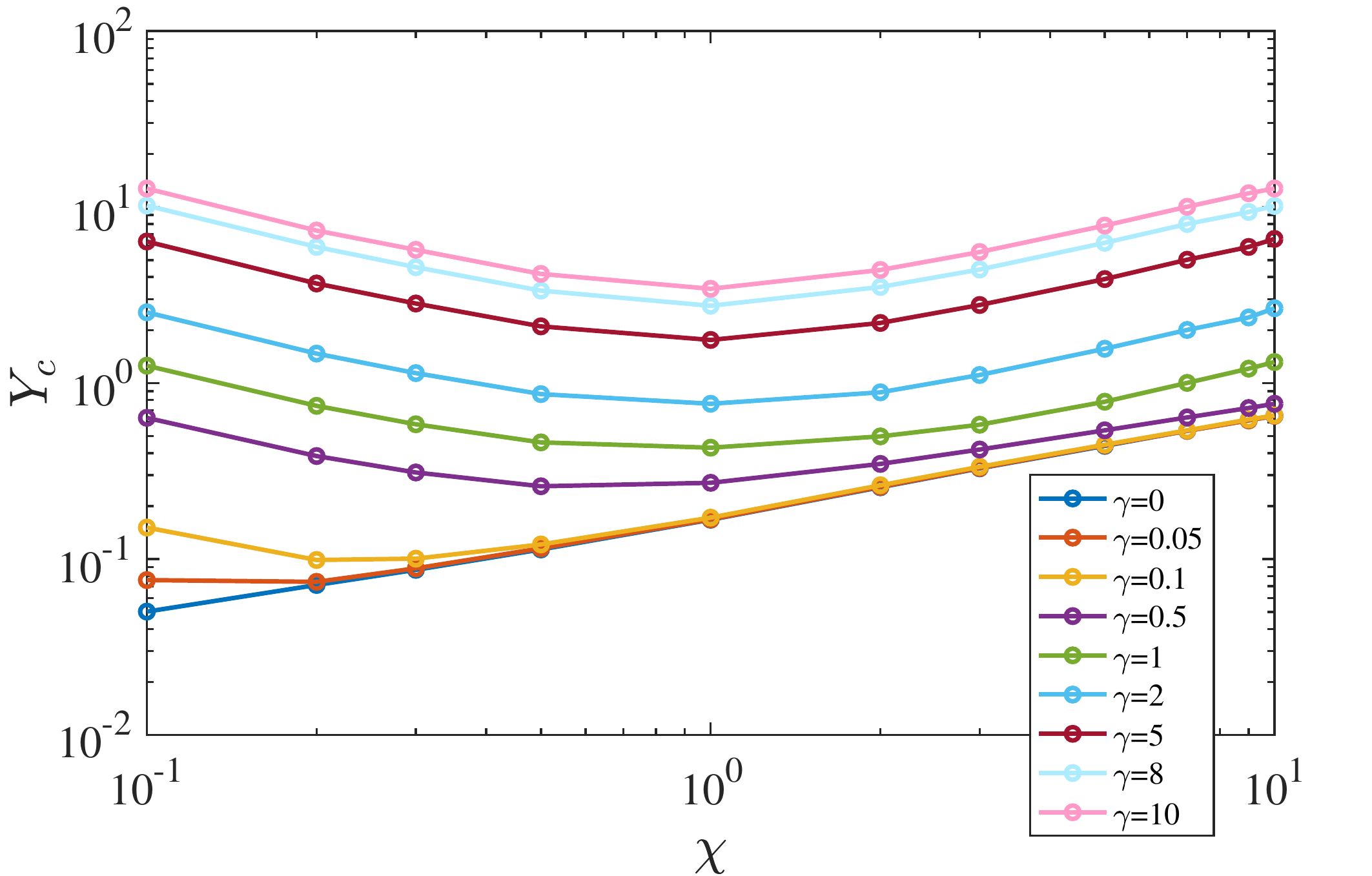}}
\caption{Critical yield number against aspect ratio for various surface tensions for quartic bubbles.}
\label{fig:Y-Chi-cyl}
\end{figure}

\begin{figure}
\centerline{\includegraphics[width=0.7\linewidth]{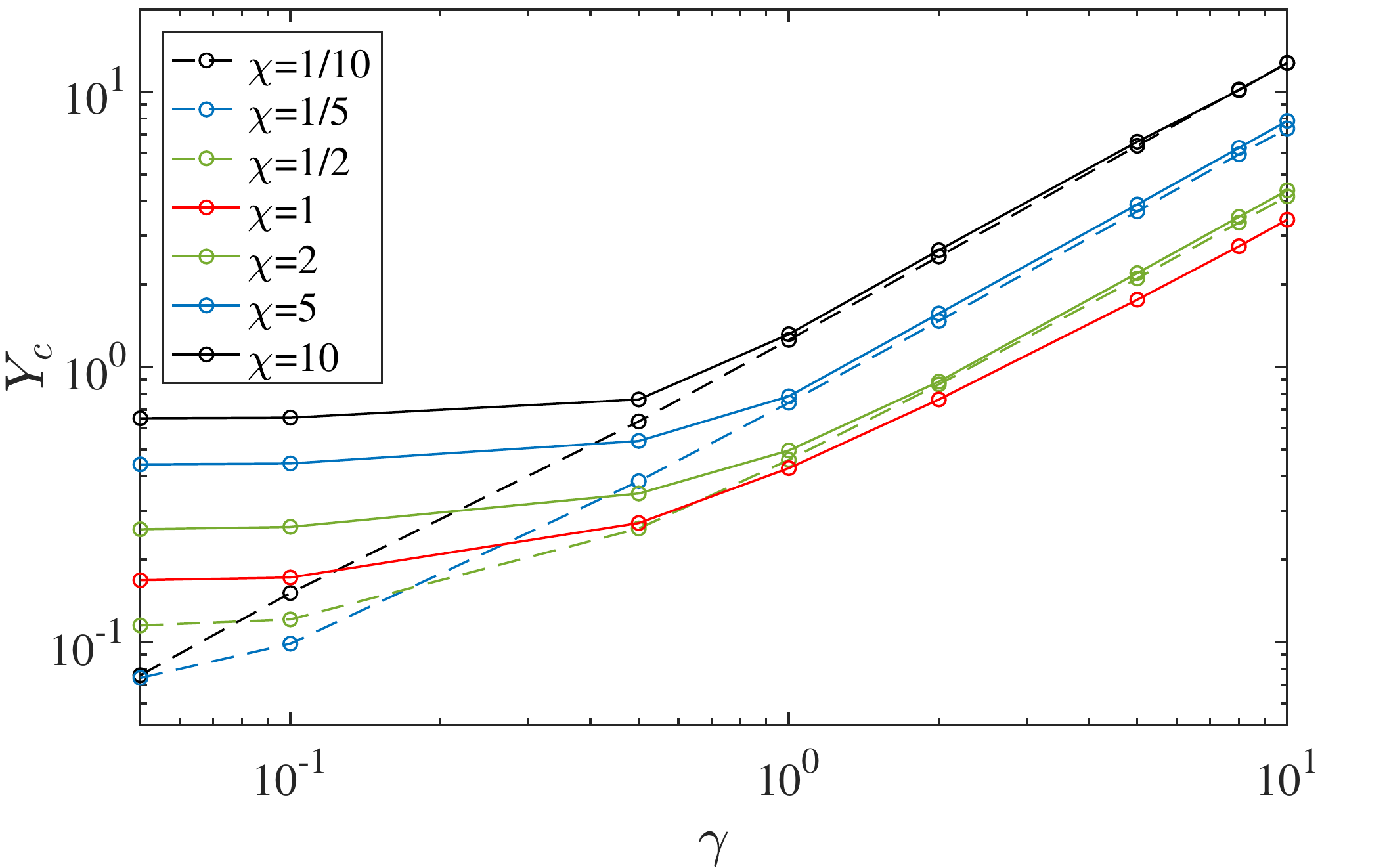}}
\caption{Critical yield number against surface tension for various aspect ratios of quartic bubbles. $\chi<1$ is plotted in dotted lines.}
\label{fig:Y-gamma-cyl}
\end{figure}

\begin{figure}
\centerline{\includegraphics[width=0.7\linewidth]{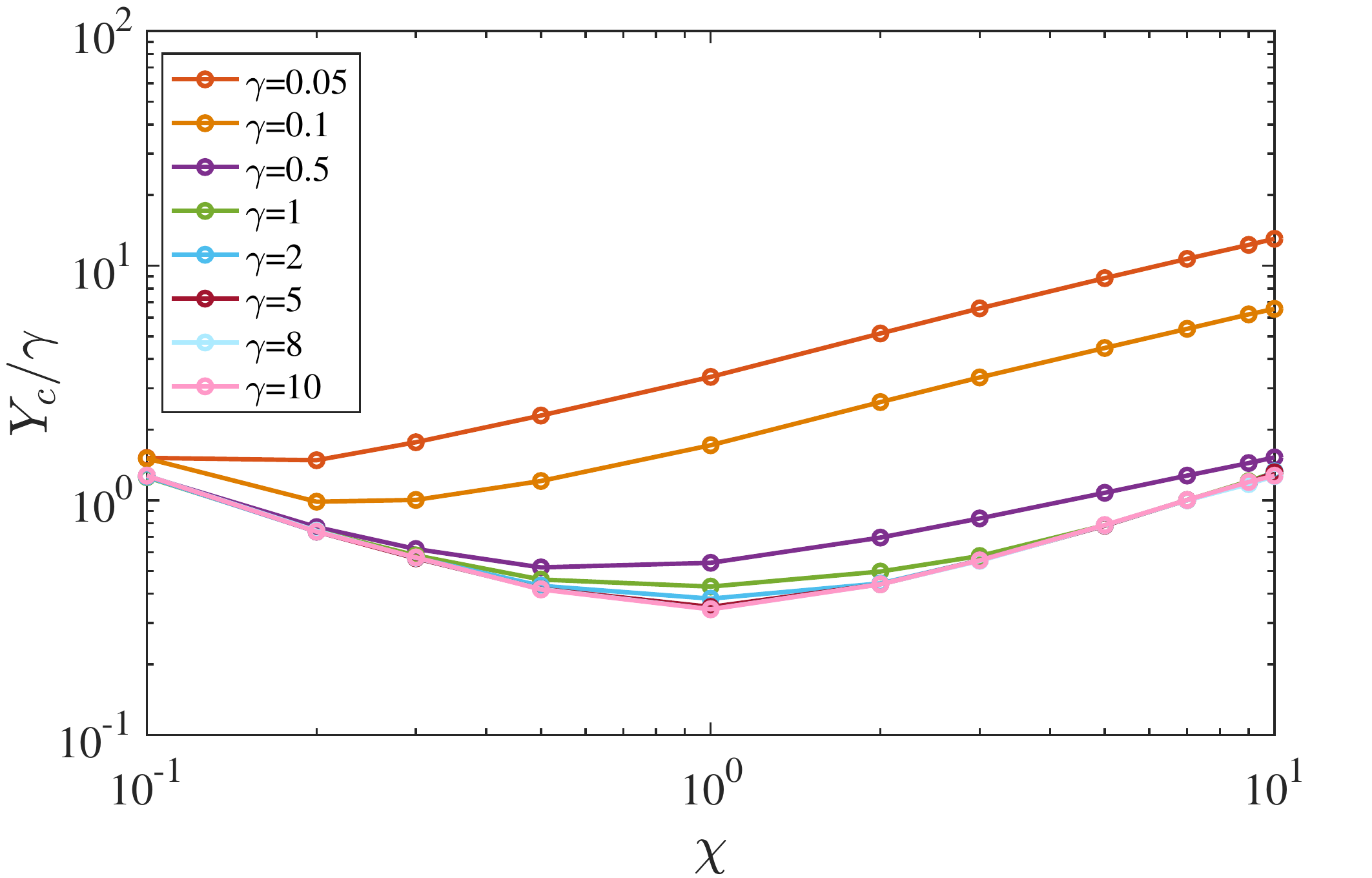}}
\caption{Yield-capillary number against aspect ratio for various surface tensions for quartic bubbles.}
\label{fig:Y_gamma-Chi-cyl}
\end{figure}

Looking at figures \ref{fig:Y-Chi-cyl} and \ref{fig:Y-gamma-cyl} we see a qualitatively similar behavior of $Y_c$. It can be observed again that higher $Y_c$ is required to stop the flow as $\gamma$ increases. For low values of $\gamma$, buoyancy dominates the flow regime and figure \ref{fig:Y-Chi-cyl} shows that the vertically oriented bubbles are more difficult to stop. Analogous to elliptical bubbles, $Y_c$ values for any specific $\chi$ and its inverse $1/\chi$ collapse on each other for high values of $\gamma$. As before, the regions of largest curvature dominate as $\gamma$ increases and orientation becomes unimportant. This observation appears to be quite general and also not confined to these two specific orientations.
Figure \ref{fig:Y_gamma-Chi-cyl} demonstrates that for high values of surface tension, the critical yield-capillary numbers tend to converge to a single curve.

Same example velocity fields around representative quartic bubbles are shown in figure \ref{fig:Chi=6,1_6}. At $\gamma = 0$ we again observe fore-aft symmetry, broken only by the surface tension. For the larger values of $\gamma$ the main velocity gradients are near the ends of bubble. Close inspection (figures \ref{fig:Chi=6,1_6}d \& h), reveals that the peak velocities are generated just off-centre-plane where curvature effects are maximal.

\begin{figure}
\centerline{\includegraphics[width=1\linewidth]{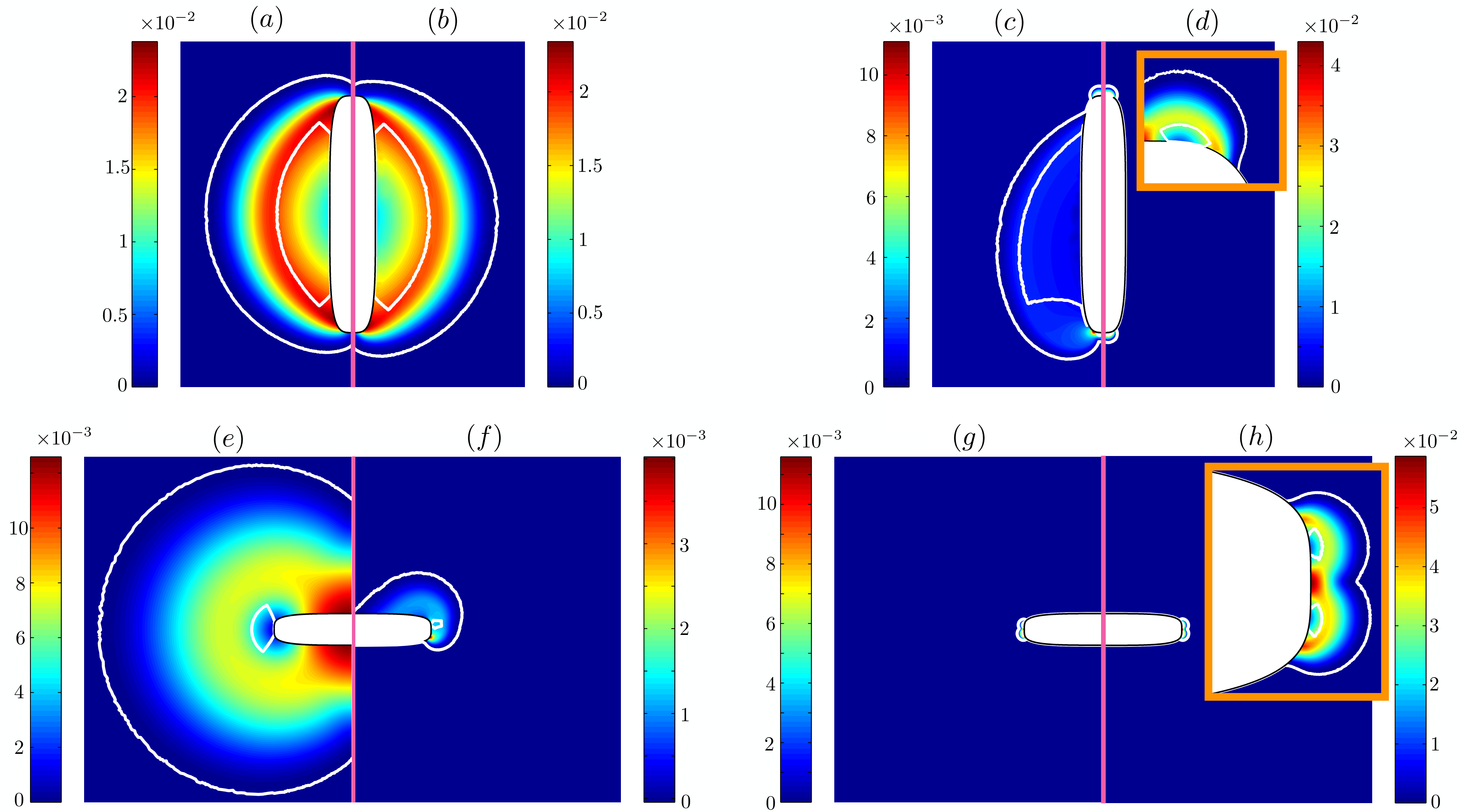}}
\caption{Velocity field around quartic bubbles for the cases of $\chi=6$ (top row) and $\chi=1/6$ (bottom row), for $\gamma$ equal to: 0 (a, e), 0.1 (b, f),  1 (c, g) and 5 (d, h). In all cases, the yield number satisfies $1-Y/Y_c =0.1$.}
\label{fig:Chi=6,1_6}
\end{figure}

\begin{figure}
\centerline{\includegraphics[width=0.7\linewidth]{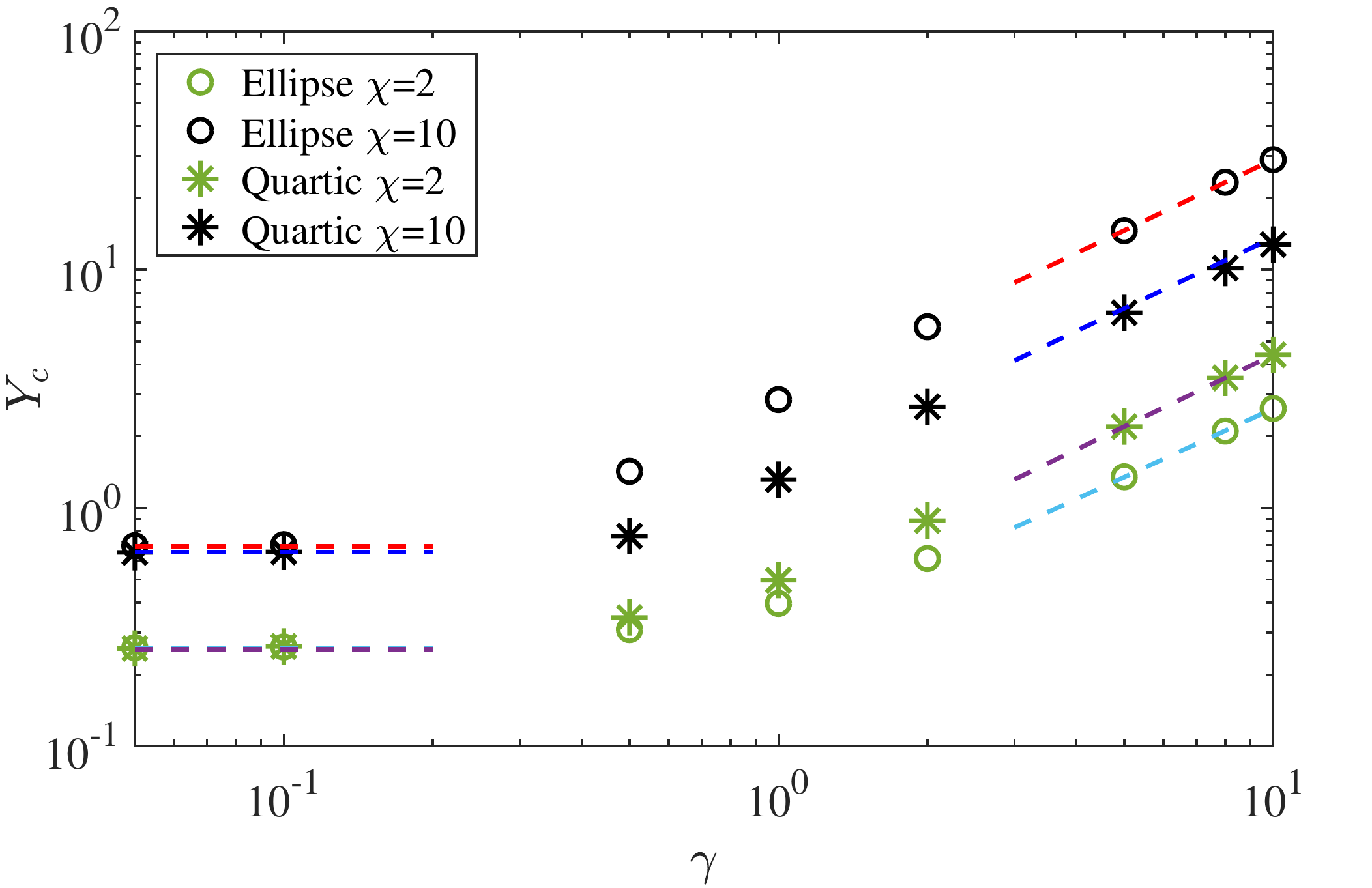}}
\caption{Behavior of the critical yield number with respect to surface tension for sample aspect ratios of $\chi=2, 10$. For high $\gamma$ this behavior is almost linear, but for low $\gamma$ it is constant.}
\label{fig:curvefit}
\end{figure}

For both the 2D elliptical and quartic bubbles the behavior of the critical yield number with $\gamma$ appears to follow the same trend, in being progressively dominated by the regions of high curvature. Looking at figure \ref{fig:curvefit}, when $\chi>1$, we find that $Y_c$ has a power-law behavior for high $\gamma$, ($Y_c=A \gamma ^B$), and is constant for small $\gamma$, ($Y_c=C$). The values of coefficients $A$ and $B$ for two sample aspect ratios of $\chi=2, 10$ are listed in Table~\ref{table:curvefit}. Evidently the exponent $B \approx 1$ as is expected from physical grounds (for large $\gamma$ when surface tension dominates buoyancy, in order to prevent motion the yield-stress must balance surface tension).

\begin{table}
\centering
\begin{tabular}{l c c c}
   & $A$ & $B$ & $C$ \\
 \hline
 Ellipse ($\chi=2$)   & 0.29 & 0.95 & 0.26 \\
 Ellipse ($\chi=10$)   & 2.98 & 0.99 & 0.69 \\
 Quartic ($\chi=2$) & 0.44 & 1.00 & 0.25 \\
 Quartic ($\chi=10$)~~~ & 1.40 & 0.96 & 0.65 \\
\end{tabular}
\caption{Example curve fit coefficients for 2D elliptical and quartic bubble shapes. $Y_c=A \gamma ^B$ for high values of $\gamma$, and  $Y_c=C$ for small values of $\gamma$.}
\label{table:curvefit}
\end{table}

\begin{figure}
\centerline{\includegraphics[width=0.8\linewidth]{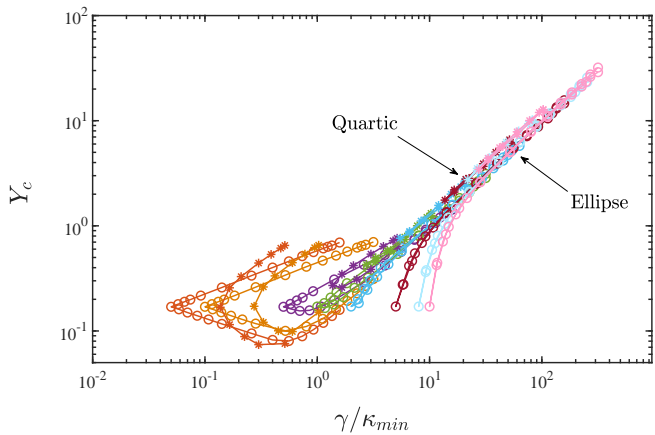}}
\caption{ $Y_c$ plotted with respect to the ratio of $\gamma$ and the minimum radius of curvature ($\kappa_{min}$) for the 2D elliptical ($\circ$) and quartic ($*$) bubbles. The color interpretation is the same as figure \ref{fig:Y_gamma-Chi-cyl}. }
\label{fig:Curvature}
\end{figure}

When $\chi<1$, $Y_c$ can be approximated with the same power-law behavior of the equivalent vertical aspect ratio (i.e.~that fitted to the data for $1/\chi$) almost over the whole range of $\gamma$. We have already seen this symmetry and data collapse in e.g.~figures \ref{fig:Y-Chi-cyl} and \ref{fig:Y-gamma-cyl}. To emphasize and confirm that the effect of curvature is the leading order, we have computed the minimal radius of curvature $\kappa_{min}$ for each shape as a function of $\chi$. We then plot $Y_c$ against $\gamma /\kappa_{min}$ for all our data, for both the ellipses and quartic bubbles; see figure \ref{fig:Curvature}. The color interpretation is the same as figure \ref{fig:Y_gamma-Chi-cyl}. Circles and stars are used for the elliptical and quartic bubbles, respectively. We observe that the data collapses onto the same curve at large $\gamma /\kappa_{min}$. There is a slight offset for the two families of bubble shapes.

\section{Axisymmetric bubbles}
\label{sec:ResultsAx}

\begin{figure}
\centerline{\includegraphics[width=1\linewidth]{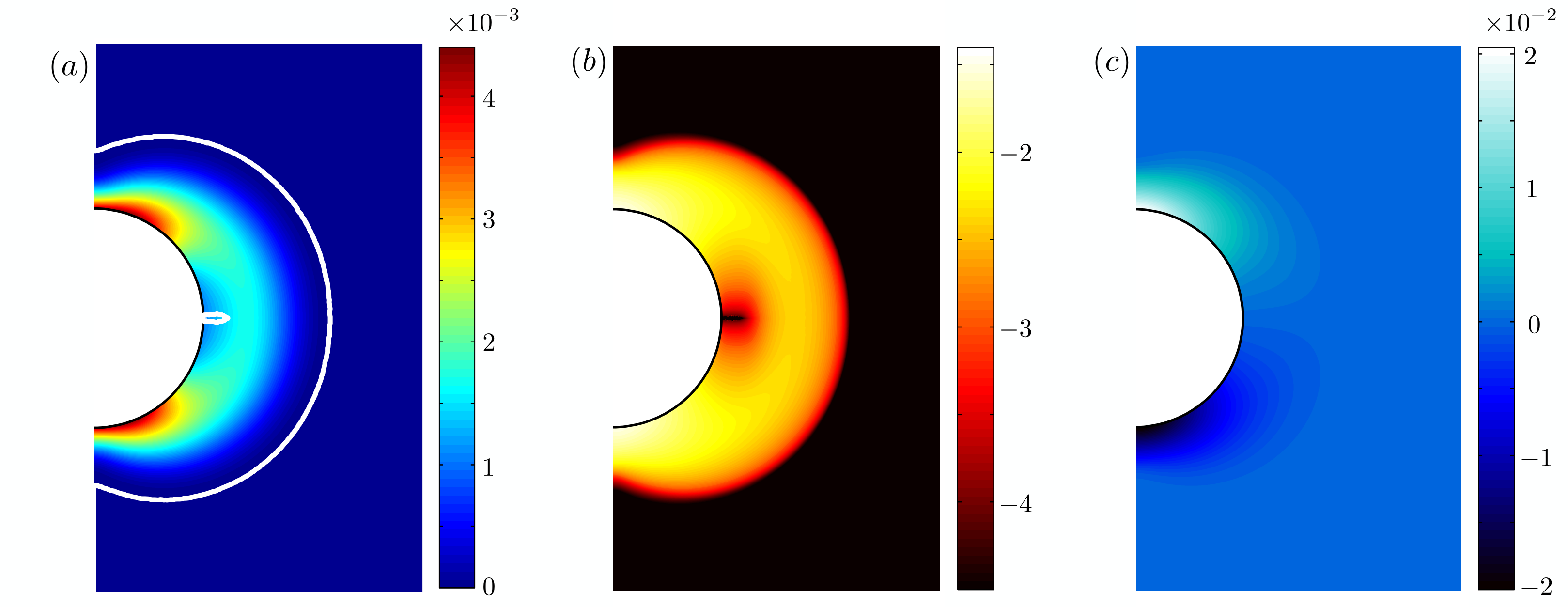}}
\caption{Axisymmetric spherical bubble at $Y=0.125$: (a) velocity field, (b) $\log (\Vert \dot{\ubgamma} \Vert)$ and (c) $\dot{\gamma}_{\theta\theta}$.}
\label{fig:Sphere}
\end{figure}

To obtain a more realistic quantitative estimate of the yield criterion, we look at similar problems but using an axisymmetric version of our codes. Figure \ref{fig:Sphere} shows the contours of the speed, the log of the rate of strain tensor and the hoop strain rate. In the axisymmetric bubbles the large kidney shaped unyielded regions attached to the bubble surface are absent due to the hoop strain rate, which we see is non-zero here, except on the equator. This gives a very small unyielded region at the equator, made possible due to the slip at the bubble surface. Note that in the case of the solid spherical particle \citep{beris1985creeping,iglesias2020computing} where there is a no-slip condition, these regions are absent (although present in 2D planar flows away from the surface).

\begin{figure}
\centerline{\includegraphics[width=0.7\linewidth]{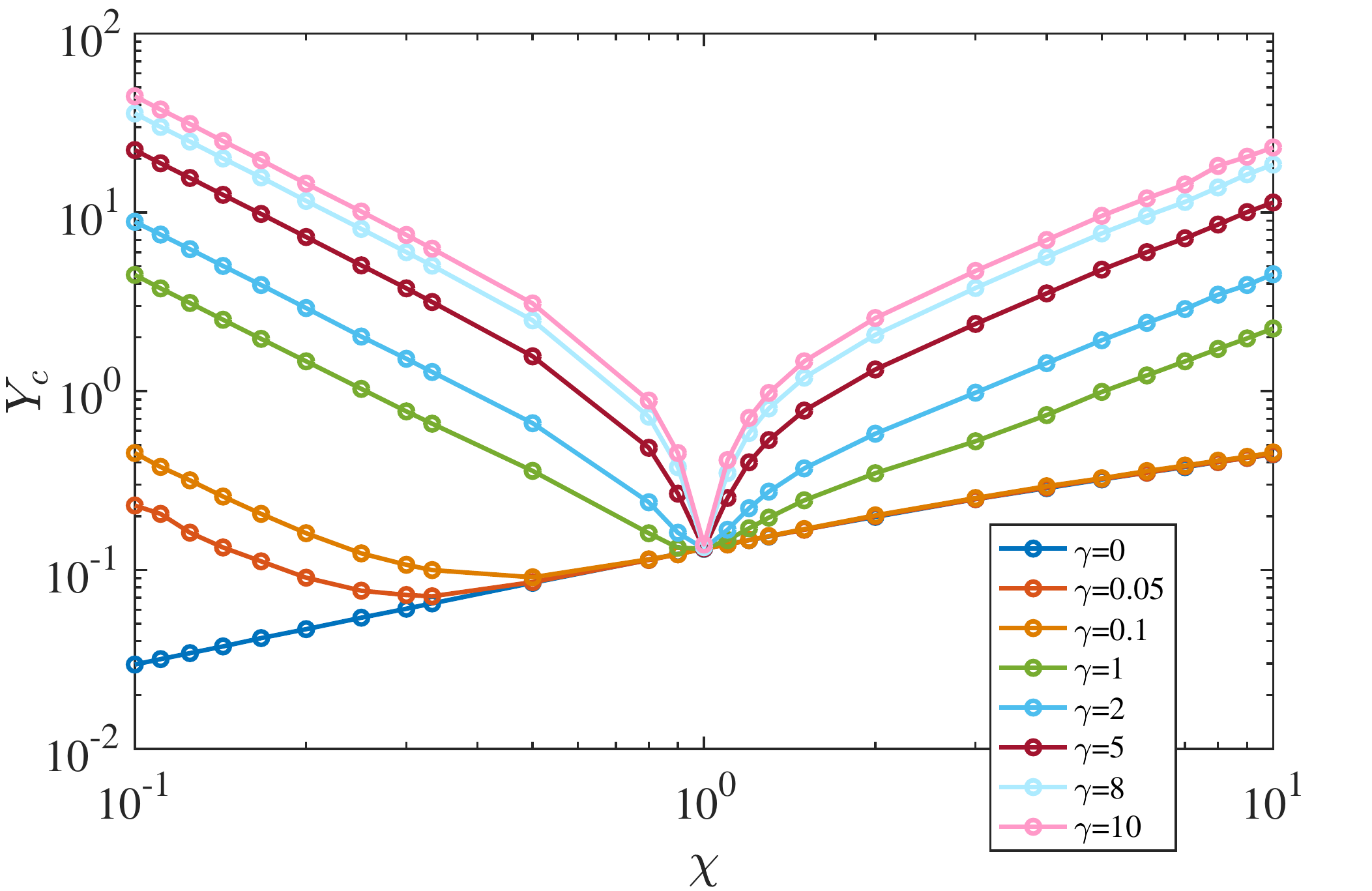}}
\caption{Critical yield number against aspect ratio for various surface tensions for ellipsoidal bubbles.}
\label{fig:Fig20}
\end{figure}

Figure \ref{fig:Fig20} plots $Y_c$ against $\chi$ for ellipsoidal bubbles. Here the ellipsoids are constructed by rotating the shape (\ref{eq:ellipse}) about the $y$-axis:
\begin{equation}\label{eq:ellipsoid}
  \frac{x^2}{a^2} +\frac{y^2}{b^2} +\frac{z^2}{a^2} =1,~~~~~a^2 b = 1;
\end{equation}
(gravity aligns with the $-y$ axis). Thus, $\chi = b/a$ has the same meaning as in 2D, viz., the bubble aspect ratio. When $\chi=1$ then $a=b=1$ which is the scaled spherical bubble radius. When $\gamma = 0$, as with the 2D bubbles,  prolate bubbles ($\chi > 1$) are harder to stop than oblate ($\chi < 1$). For the spherical bubble $Y_c$ is independent of the surface tension, as also recently stated by \citet{deoclecio2021bubble}. At first glance the ellipsoidal results qualitatively resemble those earlier in figure \ref{fig:Y-Chi} for planar bubbles.

For $\gamma = 0$, the slope of $Y_c$ in figure \ref{fig:Fig20} changes, from $\sim \chi^{2/3}$ for $\chi \ll 1$ to $\sim \chi^{0.47}$ for $\chi \gg 1$. As the bubble height now scales with $b \sim \chi^{2/3}$ the arguments for static pressure governing $Y_c$ are inadequate for prolate bubbles, and seem non-intuitive for oblate bubbles. Effectively, the flow characteristics and 3D volumetric effects must come into play. Fortunately, the functional analytic framework developed is valid and we may expect that $Y_c \approx L(\boldsymbol{u})/j(\boldsymbol{u})$ for $\boldsymbol{u}$ that are close to the solution at the yield limit. We may now construct a phenomenological explanation for the asymptotic behaviour observed. For oblate bubbles, we observe the largest velocities are directly above/below the bubbles, where fluid is pushed out of the way and this is also where significant velocity gradients are found. We therefore have $\Vert\dot{\ubgamma}\Vert \sim U_b/a $ and consequently $j(\boldsymbol{u}) \sim a^3 U_b/a = a^2 U_b$, where $U_b$ represents the mean bubble rise velocity. On the other hand, since the fluids are considered incompressible, we also have $L(\boldsymbol{u}) \sim U_b$, i.e.~what flows up must flow down, meaning that $L(\boldsymbol{u})/j(\boldsymbol{u}) \sim 1/a^2 \sim \chi^{2/3}$ as we have observed.

For $\chi > 1$ we generally find that the largest speeds are found near the 2 ends of the bubble, but decay outwards to the boundary of the yielded envelope. If we denote the outer radius of the yield envelope at $y=0$ by $R_{\perp}$ we find that $R_{\perp} \gg a$ as $\chi$ increases. All of the fluid displaced by the bubble must pass through the plane $y=0$. Thus, the fluid velocity scales as $U_b (a/R_{\perp})^2$. The yielded region around the bubble has height $\sim b$ and hence $L(\boldsymbol{u}) \sim U_b (a/R_{\perp})^2 \times [bR_{\perp}^2] \sim U_b$, as must also follow from incompressibility, (note $a^2b = 1$). On the other hand, $\Vert\dot{\ubgamma}\Vert  \sim U_b ~a^2/R_{\perp}^3$ so that $j(\boldsymbol{u}) \sim  U_b/R_{\perp}$, from which $Y_c \approx L(\boldsymbol{u})/j(\boldsymbol{u}) \sim R_{\perp}$. Examination of $R_{\perp}$, taken from our results for $\chi  > 1$ does indeed show that $R_{\perp} \sim \chi^{0.47}$. This really just establishes consistency of our results, i.e.~$Y_c \sim R_{\perp}$, but not the reason for the specific exponent $\chi^{0.47}$. In terms of the fit to the data in figure \ref{fig:Fig20}, it is worth commenting that for $\chi > 1$ the fit of the slope to $\chi^{0.47}$ is not as clean as for $\chi < 1$ with $Y_c \sim \chi^{2/3}$. Extending our computations to the range $\chi > 10$ would be helpful to better evaluate the asymptotic limit.

\begin{figure}
\centerline{\includegraphics[width=0.7\linewidth]{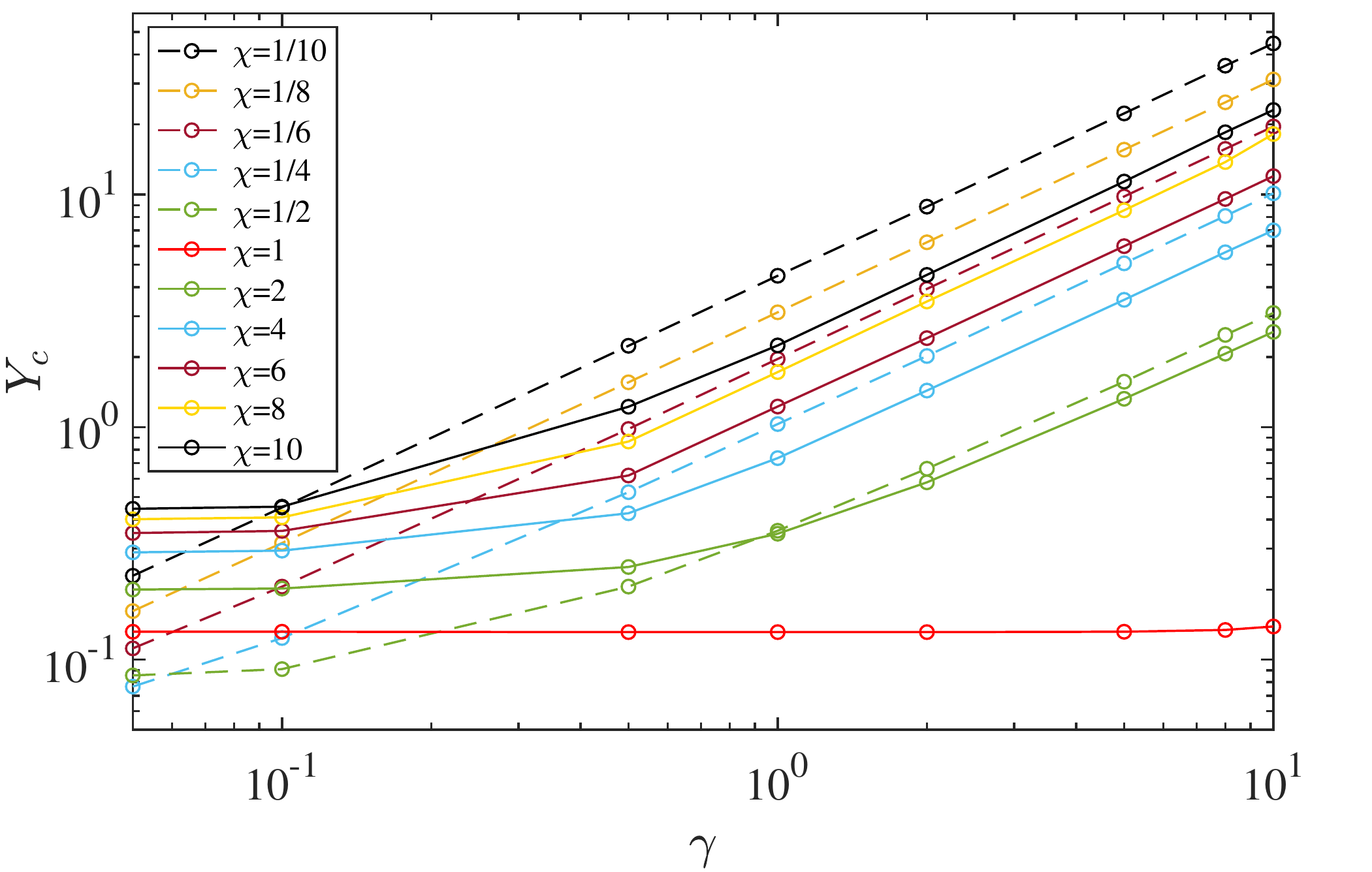}}
\caption{Critical yield number against surface tension for various aspect ratios of ellipsoidal bubbles. $\chi<1$ is plotted in dotted lines.}
\label{fig:Fig21}
\end{figure}

\begin{figure}
\centerline{\includegraphics[width=0.7\linewidth]{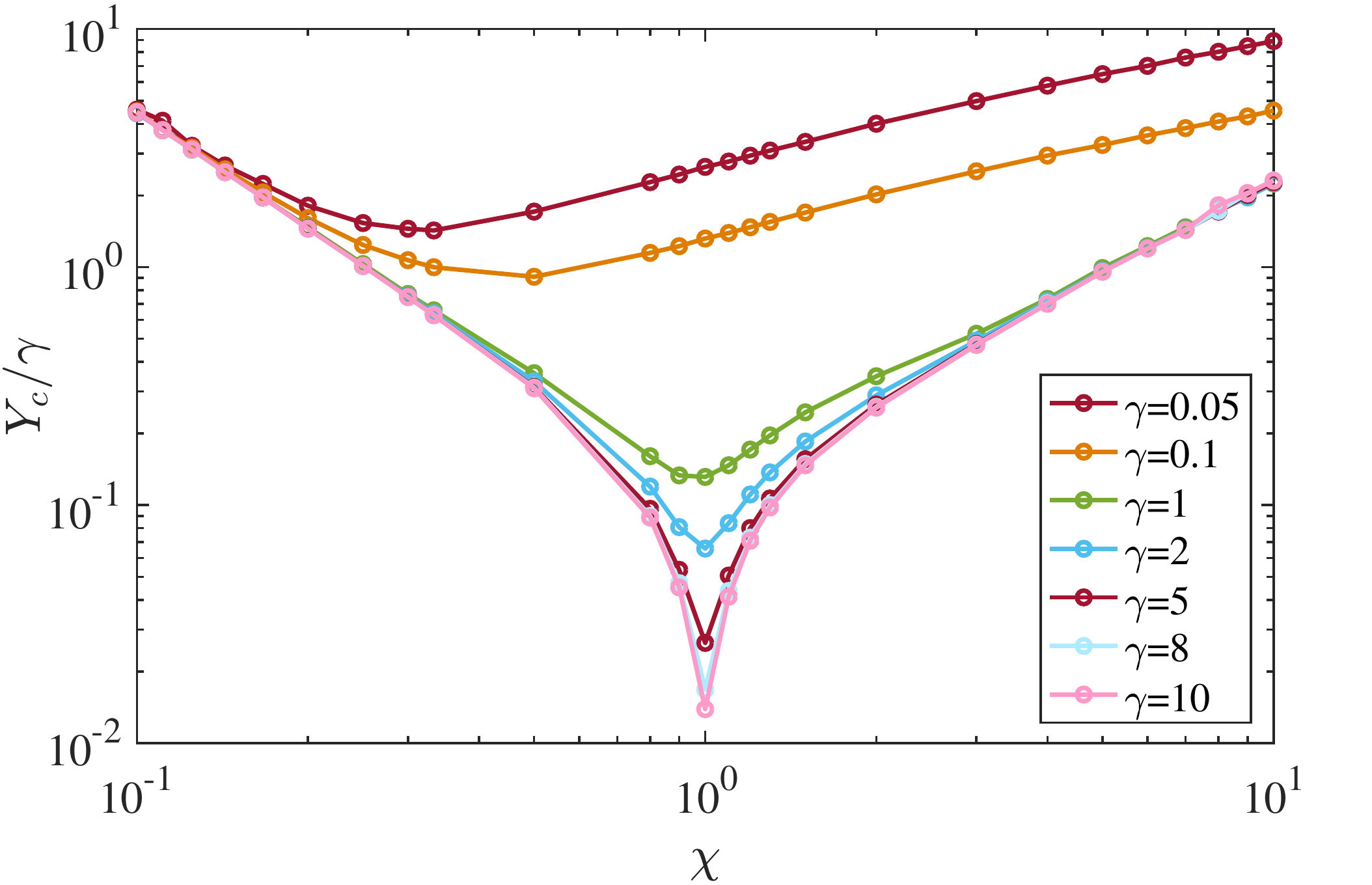}}
\caption{Yield-capillary number against aspect ratio for various surface tensions for ellipsoidal bubbles.}
\label{fig:Fig22}
\end{figure}

Looking now at $\gamma  > 0$, we see that although $Y_c$ increases for all $\chi \not=1$, unlike the planar cases the profiles of $Y_c$ do not become symmetric about $\chi = 1$, i.e.~comparing $\chi$ and $1/\chi$. This is because for the three-dimensional bubbles $\chi$ and $1/\chi$ do not yield the same shape; see equation (\ref{eq:ellipsoid}). This is seen more clearly in figure \ref{fig:Fig21} which plots $Y_c$  against $\gamma$ for different pairs of $\chi$ and $1/\chi$: the curves that do not overlap directly at large $\gamma$. The physical reasoning for the increase is the same as for the planar bubbles: at large $\gamma$ the dominant contribution to the pressure in the fluid comes from surface tension, and hence buoyancy is only a secondary influence on yielding. This is captured in figure \ref{fig:Fig22} which demonstrates that for high values of surface tension, the critical yield-capillary numbers tend to collapse onto the same limiting curve. It is also interesting to note that the $Y_c$ values for small $\chi$ are larger than those for large $\chi$, i.e.~comparing values for $\chi$ and $1/\chi$. This is a purely geometric effect of viewing the change in shape via the aspect ratio $\chi$. For the oblate ellipsoids, as $\chi \ll 1$ there is a single minimal radius of curvature that scales as $\chi^{5/3}$, so that the pressure jump at the interface is of order $\sim \gamma \chi^{-5/3}$. For the prolate ellipsoids, as $\chi \gg 1$ there are two identical minimal radii of curvature that scale as $\chi^{-4/3}$, leading to a pressure jump at the interface that is of order $\sim \gamma \chi^{4/3}$. When we compare the slopes of the curves for large $\gamma$ in figure \ref{fig:Fig20} we do indeed find $Y_c \sim \chi^{-5/3}$ and $Y_c \sim \chi^{4/3}$ for small and large $\chi$, respectively.

To allow comparison of the velocity fields with the 2D planar results, we compute the flows around the same aspect ratio ellipsoids as in figure \ref{fig:Chi=2,0.5}. Largely we see the same behavior as for the 2D planar elliptical bubbles, with the exception of the yield surfaces around the bubble equator (see figure \ref{fig:Chi=2,0.5Axi}). There also seems to be much less rigid rotational motion around the bubble surface, which is likely due to the additional hoop strain rate component.

Finally, we have also modeled bubble shapes that resemble the inverse teardrop shape often observed. We use the following basic parameterization of the surface:
\[ x=a~cos(t), y=b~sin(t)+c~( 1+cos(2t) )  ,  \]
where $t \in [-\pi/2, \pi/2]$. This shape is rotated about the $y$-axis. Adjusting the coefficients enables us to obtain various aspect ratios and the whole shape is scaled to give the correct volume. Here, we present two aspect ratios ($\chi=b/a=1~ \&~ 2$) and two surface tension values ($\gamma=0, 1$). As seen in figure \ref{fig:AxisymmetricVel}, the velocity is focused more at the tail of the bubble, and hence the yield surface is also oriented towards the tail. This is also where the radius of curvature is lowest. This behavior is most apparent in figure \ref{fig:AxisymmetricVel}d. Note that these bubbles are mobile and we are relatively far from $Y_c$. The large velocity gradients (hence stress) at the bubble tail suggest that these bubbles will deform significantly here. In other words, steady propagation of a shape such as these is not possible. Since steady bubble motion of similar shapes is actually observed experimentally (see e.g. figure 6 in \citet{pourzahedi2021eliminating}), this supports the suggestion often made, that other rheological effects are responsible for the tail.

\begin{figure}
\centerline{\includegraphics[width=1\linewidth]{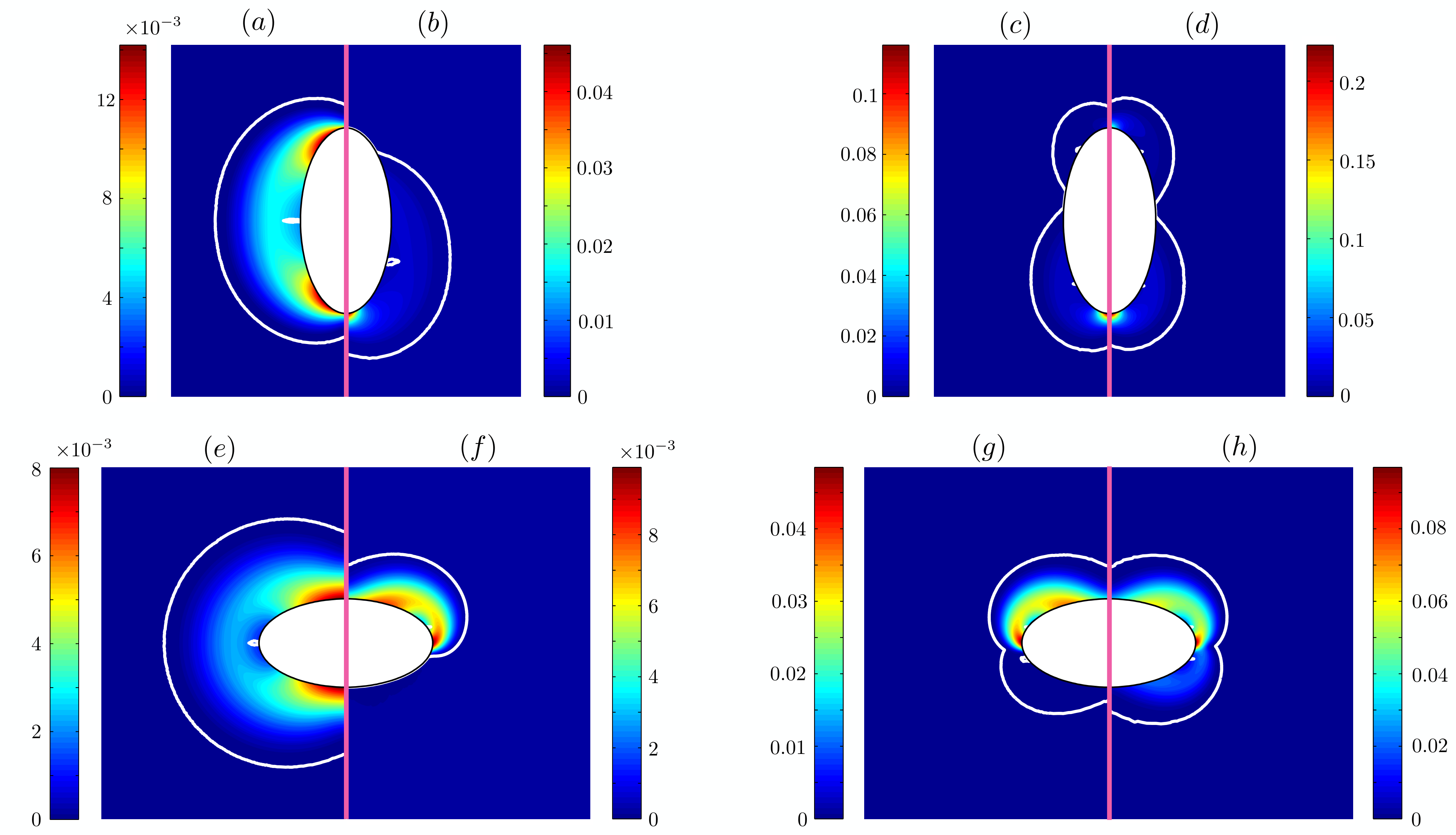}}
\caption{Velocity field around axisymmetric ellipsoidal bubbles for the cases of $\chi=2$ (top row) and $\chi=0.5$ (bottom row), for $\gamma$ equal to: 0 (a, e), 1 (b, f),  5 (c, g) and 10 (d, h). In all cases, the yield number satisfies $1-Y/Y_c =0.1$.}
\label{fig:Chi=2,0.5Axi}
\end{figure}

\begin{figure}
\centerline{\includegraphics[width=1\linewidth]{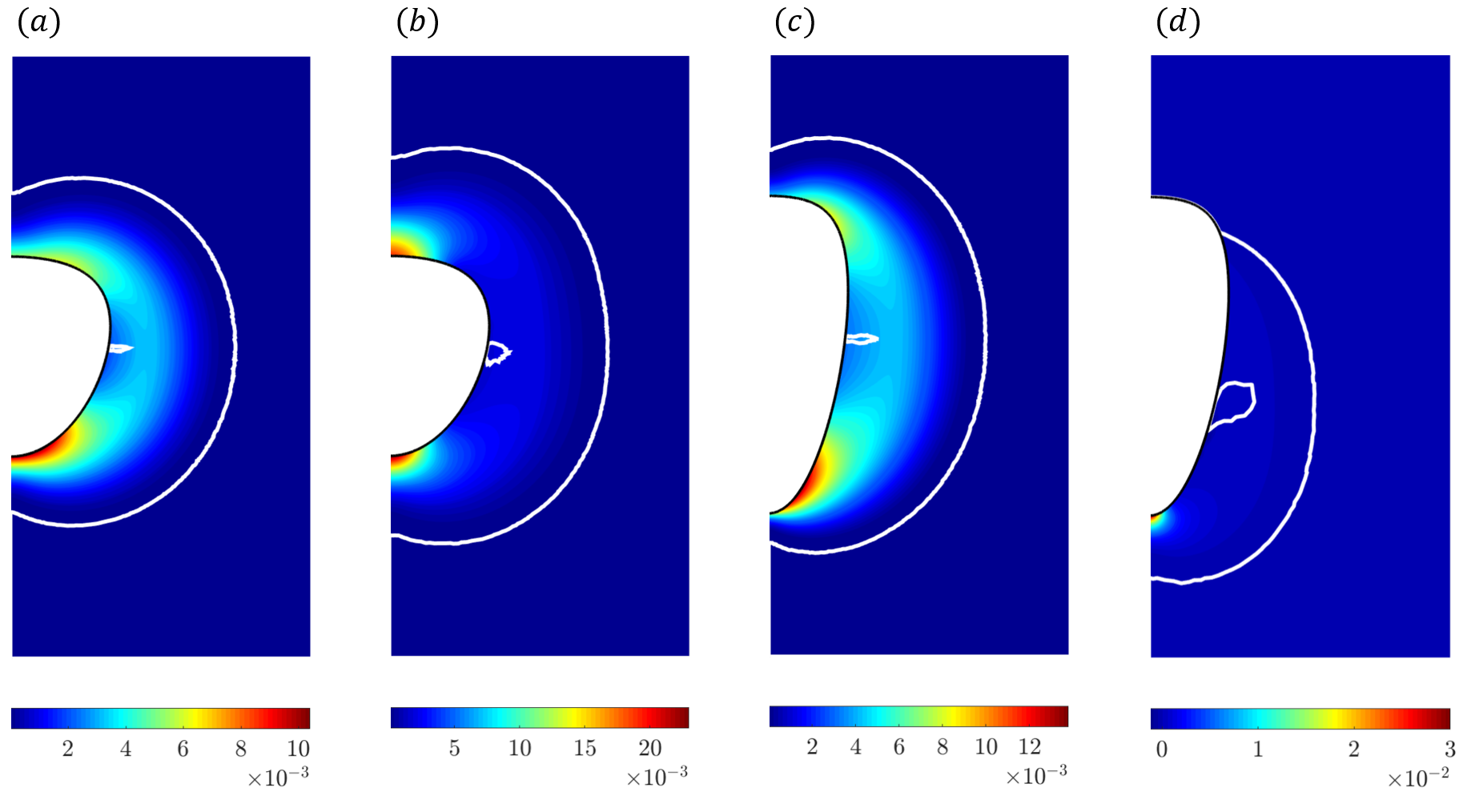}}
\caption{Velocity field around an axisymmetric teardrop bubble: (a,b) $\chi=1$ \& $c/a=0.15$, (c,d) $\chi=2$ \& $c/a=0.4$. The surface tension ($\gamma$) equals to (a,c) 0 and (b,d) 1. In all cases, the yield number satisfies $1-Y/Y_c =0.1$.}
\label{fig:AxisymmetricVel}
\end{figure}

\section{Discussion and conclusions}

This paper studied the critical yield number required to stop the motion of different bubble shapes, studying a wide range of aspect ratios and surface tension values. The computations have been conducted using both AL and FISTA algorithms, both of which are able to capture zero strain rates reliably in unyielded regions. Coupled with the adaptive meshing methods, these capture the yield surfaces well, as has been the case with many other flows. The good agreement between methods confirms the reliability of our results. Slipline theory has also been used to validate the results for 2D elliptical bubbles for the case of no surface tension, providing a close upper bound for $Y_c$.
Table~\ref{table:benchmark} presents the results of $Y_c$ for sample aspect ratios of $\chi=0.2, 1, 5$ in the absence of surface tension. For the case of a 2D circle, \citet{singh2008interacting} have reported a value of 0.167.
For the case of a sphere, \citet{tsamopoulos2008, dimakopoulos2013steady, deoclecio2021bubble} have reported values of 0.143, 0.129 and 0.133, respectively. There is good agreement for the critical yield number between previous studies and our data. Thus, the results tabulated in Table~\ref{table:benchmark} can be used as benchmarks for future computational studies. Slightly different, \citet{karapetsas2019dynamics} have found a value of 0.175 for a spherical bubble using an analytical solution based on a Lagrangian formalism.

\begin{table}
\centering
\begin{tabular}{l | c c c}
 $\chi$   & 0.2 & 1 & 5 \\
 \hline
 2D ellipse ~~   & 0.073& 0.172 & 0.460 \\
 Ellipsoid ~~~  & 0.047 & 0.132 & 0.321\\
\end{tabular}
\caption{Sample computed $Y_c$ for single 2D ellipses and axisymmetric ellipsoid shapes for the case of $\gamma=0$.}
\label{table:benchmark}
\end{table}

The shapes studied include 2D elliptic and quartic, axisymmetric ellipsoidal and inverted teardrop. There are common qualitative features of the results as the aspect ratio is varied and as the surface tension comes to dominate buoyancy. Regarding the effect of aspect ratio on the critical yield number, it was found that when the flow is buoyancy-dominated, prolate bubbles require more yield-stress to remain stationary compared to oblate bubbles. However, when the flow is surface tension-dominated, a similar yield-stress is required to stop the motion of both vertical and horizontal bubbles. The critical yield number was found to increase with surface tension. As surface tension becomes dominant, the yield capillary number is the relevant balance to study.

As approximate guidance, a $1.5$ Pa yield-stress is sufficient to hold a spherical bubble of 1 mm diameter static in an aqueous gel. A prolate bubble of the same volume with a 3:1 aspect ratio would require a $3$ Pa yield-stress in the absence of surface tension and this could be many time larger for $\gamma > 0$. Yield-stresses of waste slurries vary enormously over the range $\sim 1-10^3$ Pa, depending on many factors. Surface tension values are more constrained and small radii of curvature are generally accompanied by small volumes, i.e.~buoyancy. Thus, some care is needed in interpretation.

It is interesting to reflect on how to validate/compare our results with experiments and their broader relevance. First, it is hard to compare with many results of experiments with moving bubbles. Bubble motion in experiments usually needs some form of injection into a yield-stress fluid. Either injection \citep{sikorski2009motion,lopez2018rising, pourzahedi2021eliminating} or over-pressure \citep{zare2018onset} are common, but in both cases there is a transient motion associated with the release/pinch-off. The bubble typically then deforms quickly on release towards a steady propagation shape. Thus, we are typically far from the critical conditions. Ongoing work instead looks at using a small seed bubble in fluid within a vacuum chamber. Pressure reduction increases the volume, but the pinch-off/release/invasion steps are eliminated and viscoelastic effects do not have time to manifest in the initial shapes.

A second point of reflection is on the meaning of the results at large/small aspect ratio, where surface tension dominates. For example, with fixed yield-stress a sufficiently long/thin bubble will not be static. Does this mean it will be released? Not necessarily, as the yielded flow will deform under the action of surface tension. Where surface tension is dominant, we presume the tendency is to pull the bubble towards circular/spherical shapes, which we have seen are more stable. It is certainly conceivable that the initial deformation results in bubbles that do not rise significantly before the shape adapts. This is a limitation of our work in only calculating steady flows.

The results on the shape and aspect ratio are also interesting in the context of more complex distributions of bubbles, such as in figure \ref{fig:bubbles}, or in the industrial waste contexts that motivated this study. If the bubble release is due to a bulk effect, such as lowering of pressure or degradation of the rheology, we see that the angular and long aspect ratio bubbles are likely to deform first. With a cyclic atmospheric variation this could result in some release but also a refinement/homogenization of the shapes and sizes of the remaining retained bubble distribution. Does nature help in this way and could this cyclic procedure be imposed industrially? If the aim is to retain bubbles, use of surfactant to reduce $\gamma$ might also be an effective strategy. Indeed, the case $\gamma = 0$ appears to be the best case scenario from the bubble trapping perspective. Other effects of bubble clouds on $Y_c$ are studied in \citet{chaparian2021clouds}.

\section*{Acknowledgements}

This research has been carried out at the University of British Columbia (UBC). Financial support for the study was provided by IOSI/COSIA and NSERC (project numbers CRDPJ 537806-18 and IOSI Project 2018-10). This funding is gratefully acknowledged. The authors also express their gratitude to the University of British Columbia for financial support via the 4YF scholarship (AP). A.\ Roustaei acknowledges the financial support by Iran's national science foundation (INSF) through contract 97013654. This computational research was also partly enabled by infrastructure provided by Compute Canada/Calcul Canada (www.computecanada.ca).

\section*{Declaration of Interests}

The authors report no conflicts of interest.

\appendix

\section{Comparison of AL and FISTA methods}\label{appA}
\label{sec:AppendixA}

\begin{figure}
\centerline{\includegraphics[width=0.7\linewidth]{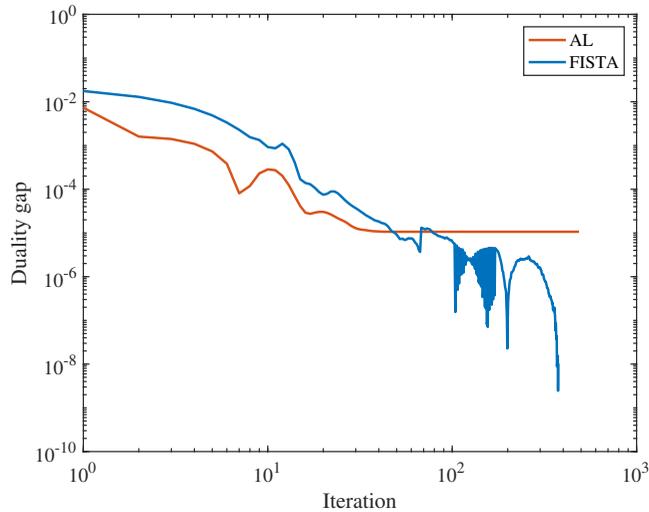}}
\caption{Comparing the duality gap of AL and FISTA methods.}
\label{fig:Dualitygap}
\end{figure}

Here we illustrate convergence behavior of the two numerical methods used in this paper (i.e.~Augmented Lagrangian and FISTA) for the case of a 2D circular bubble. The difference between the velocity minimization (primal) and stress maximization (dual) functionals, also known as the duality gap, is plotted for successive iterations in figure \ref{fig:Dualitygap}. As investigated by \cite{treskatis2018practical}, in both methods the duality gap will shrink successively but the convergence of FISTA is superior, as is clear from figure \ref{fig:Dualitygap}.

In terms of the velocity fields or yield surface positions, the differences are not discernible. Both algorithms are operating with the exact Bingham model, as opposed to a regularization approach, and thus approximate the rigid unyielded regions correctly, which is the main point for computations that explore stopping/static flows. Practically speaking, in terms of computing the static limits and yield surface shapes, as here, more relevant than the (FISTA or AL) algorithm is the number of the mesh adaptations used.

\bibliographystyle{jfm}

\bibliography{viscoplastic}


%
%
%
%
%
%
%


\end{document}